\documentclass[doublecolumn,9pt,double,romanappendices]{IEEEtran}
\usepackage[cmex10]{amsmath}
\usepackage{graphics}
\usepackage{mdwmath,mdwtab,url}
\usepackage[caption=false,font=footnotesize]{subfig}
\usepackage{stfloats}
\usepackage{nomencl}
\usepackage{amssymb}
\usepackage{rotating,geometry}
 \usepackage{pxfonts}
\usepackage{psfrag}
\usepackage{color}
\usepackage{hyperref}
\usepackage{cite,balance}
\usepackage{epstopdf}
\usepackage{mathtools,floatrow }

\newtheorem{theorem}{ {Theorem}}

\newtheorem{definition}{{Definition}}
\newtheorem{property}{\ti{Property}}
\newtheorem{lemma}{ {Lemma}}
\newtheorem{remark}{ {Remark}}

\newcommand{\mb}{\mathbb}
\newcommand{\dv}{\textbf} % vector in math bold font
\newcommand{\mc}{\mathcal} % variable in math calligraphic font
\newcommand{\ti}{\textit}
\newcommand{\tf}{\mathbf}

\definecolor{brown}{rgb}{.75,.5,.5}
\begin{document}
%
%iffalse
\title{On SDoF of Multi-Receiver Wiretap Channel With Alternating CSIT}
\author{Zohaib~Hassan~Awan, Abdellatif~Zaidi, and Aydin Sezgin

\thanks{Copyright (c) 2013 IEEE. Personal use of this material is permitted. However, permission to use this material for any other purposes must be obtained from the IEEE by sending a request to pubs-permissions@ieee.org}  

\thanks{Zohaib Hassan Awan and Aydin Sezgin are with Institute of  Digital Communication Systems, Ruhr-Universit\"{a}t Bochum, 44780 Bochum, Germany. Email: \{zohaib.awan, aydin.sezgin\}@rub.de}

\thanks{Abdellatif Zaidi is with Universit\'e Paris-Est Marne-la-Vall\'ee, 77454 Marne-la-Vall\'ee Cedex 2, France. Email: abdellatif.zaidi@univ-mlv.fr}

 \thanks{This work is supported by the German Research Foundation, Deutsche Forschungsgemeinschaft (DFG), Germany, under grant SE 1697/11. }

\thanks{ The result in this work was presented in part at the IEEE International Symposium on Information Theory, Honolulu, USA, Jun.-Jul. 2014~\cite{isit14}.}
 
}

\maketitle
 
\begin{abstract}
 We study the problem of secure transmission over a  Gaussian  multi-input single-output (MISO) two receiver channel with an external eavesdropper, under the assumption that the state of the channel which is available to each receiver is conveyed either perfectly ($P$) or with delay ($D$) to the transmitter. Denoting by $S_1$, $S_2$, and $S_3$ the channel state information at the transmitter (CSIT) of user 1, user 2, and eavesdropper, respectively, the overall CSIT can then alternate between eight possible states, i.e., $(S_1,S_2,S_3) \in \{P,D\}^3$. We denote by $\lambda_{S_1 S_2 S_3}$ the fraction of time during which the state $S_1S_2S_3$ occurs. Under these assumptions, we first consider the Gaussian MISO wiretap channel and characterize the secure degrees of freedom (SDoF). Next, we consider the general multi-receiver setup and characterize the SDoF region of fixed hybrid states $PPD$, $PDP$, and $DDP$. We then focus our attention on the symmetric case in which 
$\lambda_{PDD}=\lambda_{DPD}$. For this case, we establish bounds on SDoF region.
The analysis reveals that alternating CSIT allows synergistic gains in terms of SDoF; and, shows that by opposition to encoding separately over different states, joint encoding across the states enables strictly better secure rates.  Furthermore, we specialize our results for the two receivers channel with an external eavesdropper to the two-user broadcast channel. We show that the synergistic gains in terms of SDoF by alternating CSIT is not restricted to multi-receiver wiretap channels; and, can also be harnessed under broadcast setting. 
\end{abstract}
 
  \section{Introduction}\label{secI}
In cellular networks, multiple nodes communicate with each-other over a shared wireless medium. Due to the broadcast and superposition nature of the wireless medium, simultaneous transmission of information over this channel emanates an important issue of interference in networks. As the communication network grows, and since due to scarcity of available resources for example, radio spectrum and available power,  the detrimental effect of interference is unavoidable. A key resource that helps mitigating the effect of interference more efficiently is the availability of CSIT. In the literature, different multi-user networks are studied under ideal assumption of perfect CSIT in~\cite{J10} (and references therein), where quality of CSIT plays a major role in aligning or canceling interference in networks. Recently, a growing body of research has attracted attention to study a wide variety of two-user CSIT models, e.g., with strictly causal (delayed) CSI in~\cite{M-AT12,vaze_broadcast}, no CSIT in~\cite{vaze_no} and with mixed CSIT (perfect delayed CSI along with imperfect instantaneous CSI) in~\cite{sheng_mix}, all from degrees of freedom (DoF) perspective. In all these models, it is assumed that \textit{symmetric} CSI is available at the transmitter, i.e., either perfect, delayed or no CSI is conveyed by \ti{both} receivers. In~\cite{Tandon-BC-partial}, Tandon \ti{et al.} studied a two-user broadcast channel with \textit{asymmetric} CSI conveyed to the transmitter. In this model, the channel to one receiver is  available instantaneously at the transmitter, while the channel to the other receiver is conveyed with some delay. The authors refer to this model as being one with partially perfect CSIT and they characterize the DoF region. Due to the random fluctuations in the wireless medium, it becomes difficult for the receivers to convey the same quality of CSI over time. In another related work~\cite{TJSP13}, Tandon \ti{et al.} studied the two-user broadcast channel by taking time varying nature of CSIT into account. Among other constraints, the authors assumed that the CSI conveyed by \ti{both} receivers can vary over time, and each receiver is allowed to convey either perfect, delayed or no CSIT to the transmitter in an \textit{asymmetric} manner. For this channel model, they characterize the full DoF region.

As said before, in wireless networks, due to the broadcast nature of the medium, information exchange between two communicating parties can be overheard by other nodes of the network for free. The adversaries (eavesdroppers) listen to this communication and can try to extract some useful information from it. In his seminal work~\cite{wyner}, Wyner introduced a basic information-theoretic model to study secrecy by taking physical layer attributes of the channel into account. For the degraded wiretap channel, in which the channel from the source-to-legitimate receiver is stronger than the one from source-to-eavesdropper, secrecy capacity is established. The wiretap channel introduced by Wyner is extended to study a variety of multi-user networks, for example,  the broadcast channel~\cite{csiszar,yanling},  multi-access  channel~\cite{tekin,liangpoor,MAC_ieee},  relay channel~\cite{lai,Z_relay_ieee},  interference channel~\cite{onur_IFC,LYT08}, and  multi-antenna channel~\cite{khisti}.  Characterizing the secrecy capacity region of these models fully can be very challenging in general. At high signal-to-noise ratio (SNR), similar to DoF, the notion of SDoF captures the asymptotic behaviour of the data rates that are allowed securely. More specifically, it shows how the secrecy capacity prelog or spatial multiplexing gain scales asymptotically with logarithm of SNR. In~\cite{khisti}, Khisti \textit{et al.} study a Gaussian multi-input multi-output (MIMO) wiretap channel in which perfect CSI of the legitimate receiver and eavesdropper is available at the transmitter; and establish the secrecy capacity as well as the SDoF. In \cite{ruoheng_MIMO_BC}, Liu \textit{et al.} generalize the model in~\cite{khisti} to the broadcast setting and characterize the secrecy capacity region. For the two-user (2,1,1)--MISO broadcast channel the optimal sum SDoF is 2, and is obtained by zero-forcing the confidential messages at the unintended receivers. Yang \textit{et al.} in~\cite{YKPS11} study the MIMO broadcast channel and show that strictly causal (delayed) CSI   is still useful from SDoF perspective in the sense that it enlarges the secrecy region, in comparison with the same setting but with no CSIT. In~\cite{Sdof-x}, Zaidi \textit{et al.} study the MIMO X-channel with asymmetric feedback and delayed CSIT and characterize  the corresponding SDoF region. Recently, in\cite{geraci,wang}, the authors studied the secure broadcast setting with mixed CSIT. Despite of all these important recent advances for models with delayed CSI at transmitters, settings in which CSI from receivers are observed with different delays are still not fully understood.  The channel that we study in this work can be seen as a step further towards better understanding this type of models.

\begin{figure}
    \centering
    \begin{floatrow}
      \ffigbox[\FBwidth]{\caption{(3,1,1,1)--Multi-receiver wiretap channel with alternating CSIT, and security constraints.}\label{model-m}}{%
          \includegraphics[scale=0.28 ]{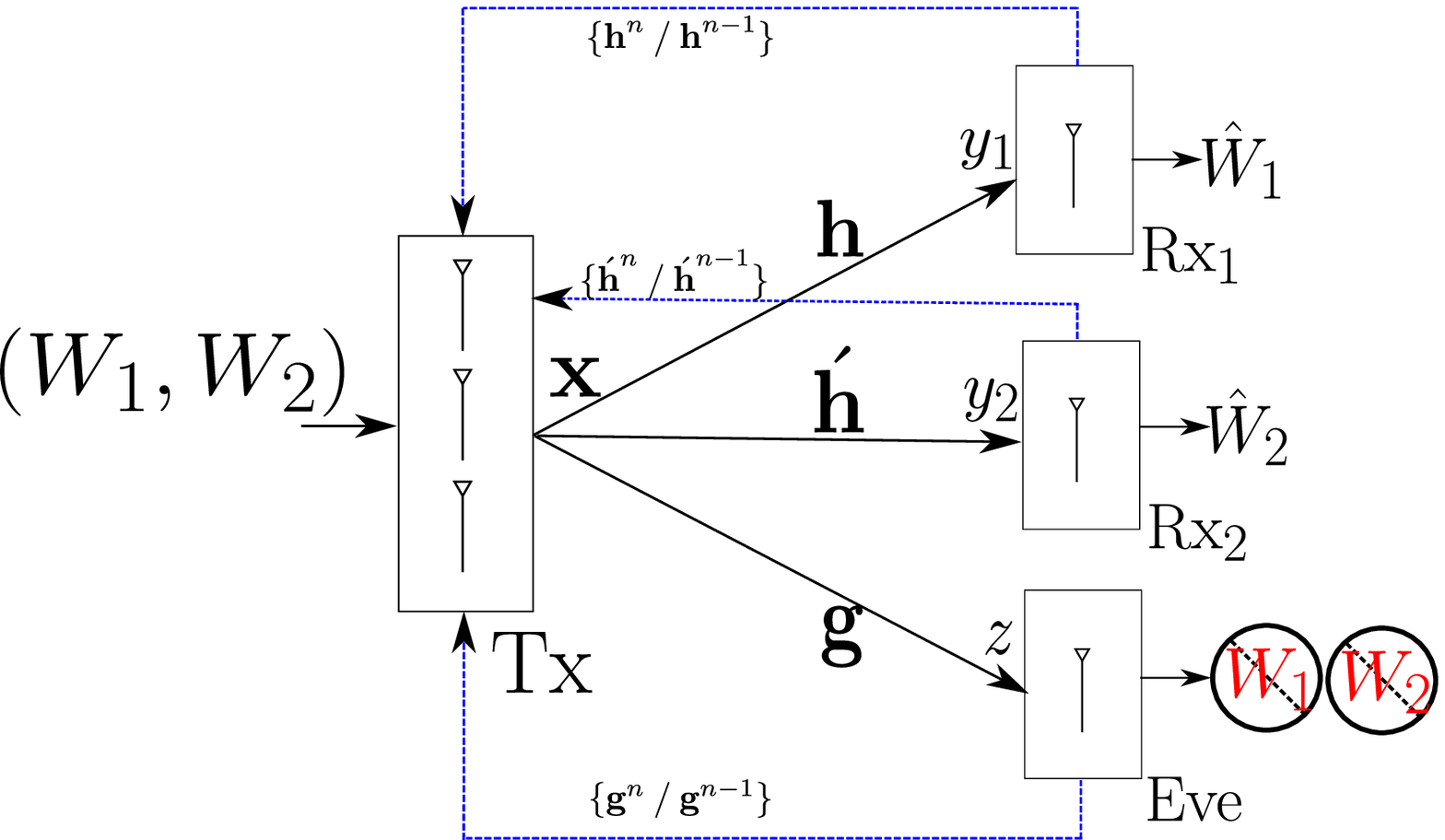}   % Just a dummy. Replace with your figure.
      }
      \ffigbox[\FBwidth]{\caption{{(2,1,1)--Two-user MISO broadcast channel with alternating CSIT, and security constraints.}}\label{model}}{%
              \hspace{-1.5em}\includegraphics[scale=.28]{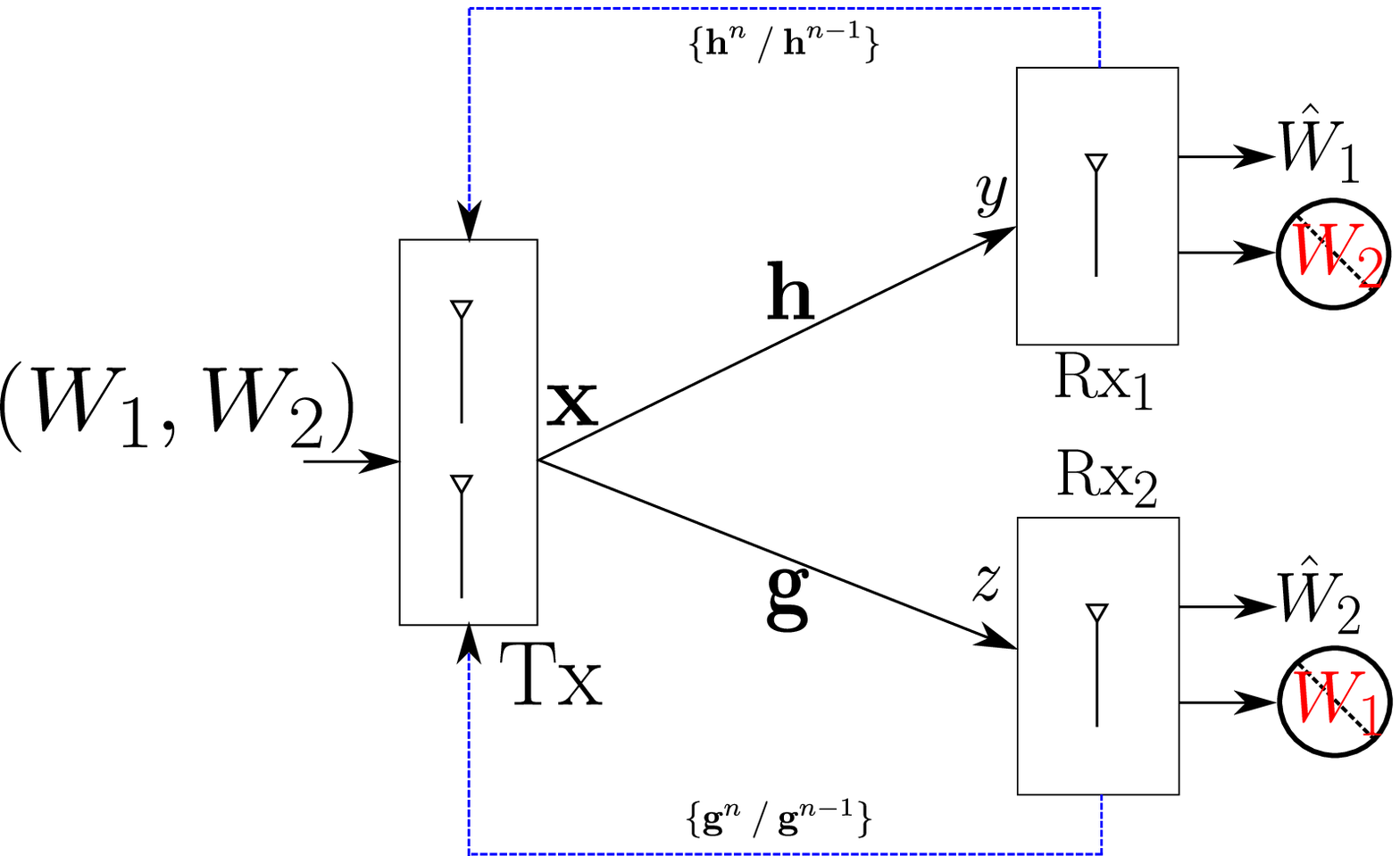} % Just a dummy. Replace with your figure.
      }
    \end{floatrow}
  \end{figure}

In this work, we consider a two-receiver Gaussian MISO channel with an external eavesdropper in which the transmitter is equipped with three antennas, and each of the three receivers is equipped with a single antenna as shown in Figure~\ref{model-m}. The transmitter wants to reliably transmit messages $W_{1}$ and $W_{2}$ to receiver 1 and receiver 2. In investigating this model we make three assumptions, namely, 1) the communication is subjected to a fast fading environment, 2) each receiver knows the perfect instantaneous CSI and also the CSI of the other receiver with a unit delay, and 3) the channel to each receiver is conveyed either instantaneously ($P$) or with a unit delay ($D$) to the transmitter. In both cases, it is assumed that the CSI is perfect. We assume that the eavesdropper is the part of the communication system, and in its desire to learn the information, is willing to convey its own CSI to the transmitter. Thus, the CSIT vector that is gotten at the transmitter from the two receivers and the eavesdropper can alternate among eight possible states, $PPP, PPD, PDP, PDD, DPP, DPD, DDP$, and $DDD$.  Furthermore, the transmitter wants to conceal the message $W_1$ that is intended to receiver 1, and the message $W_{2}$ that is intended to receiver 2 from the external eavesdropper.  We assume that the eavesdropper is passive, i.e., it is not allowed to modify the communication. The model that we study can be seen as similar to the one in~\cite{mu-delay} but with alternating CSIT setting. 
We consider the case of perfect secrecy and focus on the asymptotic behavior of this model, where system performance is measured by SDoF.
 
The main contributions of this work are summarized as follows. We first consider a (3,1,1)-MISO wiretap channel with alternating CSIT, and characterize fully the optimal SDoF for this model. The coding scheme in this case is based on an appropriate combination of schemes that we develop for fixed CSIT configurations, namely, $PP$, $PD$, $DP$ states and the one that is developed previously for $DD$ state in~\cite{YKPS11}. The converse proof follows by extending the proof of~\cite{YKPS11} developed in the context of  wiretap channel with delayed CSIT to the case with alternating CSIT; and, also, uses some elements from the converse proof of~\cite{TJSP13} established for the broadcast model with alternating CSIT by taking imposed security constraints into account. We note that, our result for the MISO wiretap model is not restricted to \textit{symmetric} case, i.e., $\lambda_{S_1 S_3}=\lambda_{S_3 S_1}$ and holds in general.

Next, we consider the multi-receiver wiretap channel as shown in Figure~\ref{model-m} and establish bounds on SDoF region. In particular, we first consider the hybrid states, $PPD$, $PDP$ ($DPP$), and $DDP$ and characterize the complete SDoF region. Afterwards, we consider the case in which the transmitter is allowed to alternate between two states, i.e.,  $PDD$ and $DPD$ equal fractions of communication time. For this case, we establish both inner and outer bounds on SDoF region. The coding scheme that we use to establish the inner bound, sheds light on how to multicast  common information securely to both receivers.  Although non-optimal in general, the results of this work show that, for the multi-receiver wiretap channel that we study, alternating CSIT not only enables interesting synergistic gains in terms of degrees of freedom in the case without secrecy constraints as was shown in~\cite{TJSP13}, but also if secrecy constraints are imposed on the communication.

Next, we specialize our results developed for the multi-receiver wiretap channel to the two-user broadcast setting. The two-user Gaussian MISO broadcast channel consists of a transmitter and two receivers,  where the transmitter is equipped with two antennas, and each of the two receivers is equipped with a single antenna as shown in Figure~\ref{model}. The transmitter wants to reliably transmit messages $W_{1}$ and $W_{2}$ to the receiver 1 and the receiver 2, respectively. Similar to the previous setup, we assume that the channel to each receiver is conveyed either instantaneously ($P$) or with a unit delay ($D$) to the transmitter. Thus, the CSIT vector that is gotten at the transmitter from the two receivers can alternate among four possible states, $PP$, $PD$, $DP$ and $DD$.  Furthermore, the transmitter wants to conceal the message $W_1$ that is intended to receiver 1 from receiver 2; and the message $W_{2}$ that is intended to receiver 2 from receiver 1. Thus, each receiver plays two different roles, being at the same time a legitimate receiver of the message that is destined to it, and an eavesdropper of the message that is destined to the other receiver.  We establish inner and outer bounds on the SDoF region of this model. As part of the main ingredients that we employ for proof of the inner bound, we develop some elementary coding schemes that can be seen as an appropriate generalization of those in~\cite{TJSP13}, tuned carefully so as to account for the imposed secrecy constraints.  The proof of our outer bound follows by carefully extending our proof for the MISO wiretap model to the broadcast setting. The outer and inner bounds that we have constructed do not agree in general; however, for the special case in which perfect CSI or strictly causal CSI is conveyed by \ti{both} receivers we recover the SDoF region in~\cite{ruoheng_MIMO_BC} and~\cite{YKPS11}, respectively.

 We now highlight the key differences between some of the results in this paper and  
a similar work which was independently done in parallel in~\cite{MTU14}. In~\cite{MTU14}, the authors studied a MISO broadcast channel with confidential messages in which the transmitter is allowed to alternate between two states, i.e., $PD$ and $DP$, equal fractions of communication time. The authors characterize the complete SDoF region. As opposed to the model in~\cite{MTU14}, the model that we study in this paper as shown in Figure~\ref{model} is more general, since it allows more leverage to the transmitter to choose between four possible states, i.e.,  $PP$, $PD$, $DP$, and $DD$. Specializing the results in this work to the model studied in~\cite{MTU14} reveals that the encoding scheme in~\cite{MTU14} outperforms the scheme that we developed in~\cite{isit14} (SDoF of $3/2$ vs. $4/3$). However, the outer bound that we have established in this work is more general; and, it subsumes the outer bound in~\cite{MTU14} and the one developed in the context of broadcast channel with delayed CSIT in~\cite{YKPS11}. Recently, in~\cite{PRS-15} the authors extended their model in~\cite{MTU14} to a more general setup in which the two receivers are allowed to convey either perfect $(P)$, delayed $(D)$ or no CSIT $(N)$ and characterized the full SDoF region. Specializing the outer bound established in~\cite{PRS-15} to our setup in Figure~\ref{model} shows that the outer bound that we establish in this work coincides with the one in~\cite{PRS-15}. Finally, it is worth noting that none of the works in~\cite{MTU14} and~\cite{PRS-15} have investigated the multi-receiver wiretap channel in Figure~\ref{model-m} that we study. The results in this paper can serve as a
stepping stone towards understanding the general class of $K$-user models.   
 
We structure this paper as follows. Section~\ref{miso-mu-sec-sys} provides a formal description of the channel model that we study along with some useful definitions. Section~\ref{secIII} states the SDoF of the $(3,1,1)$--MISO wiretap channel.
In Section~\ref{fixed}, we study the multi-receiver wiretap channel with fixed hybrid states; and, in Section~\ref{alt} we extend our results for this model  to the alternating CSIT setting. Section~\ref{secIV}, provides the description of the  two-user broadcast channel and states the main results. The formal proof of the coding scheme that we use to establish the inner bound for the two-user MISO broadcast channel is given in Section~\ref{schemes}. Finally, in Section~\ref{conclusion} we conclude this paper by summarizing its contributions.
 
 \vspace{0.5em}
\textit{ {Notations:}}
We will use the following notations throughout this work. Boldface upper case letter $\tf X$ denotes matrices, boldface lower case letter $\tf x$ denotes vectors, and calligraphic letter $\mc X$ designates alphabets; at each time instant $t$, $\tf x_t$ denotes $[x_{t1},\hdots,x_{tn}]$. For integers $i \leq j$, $\tf X^{j}_{i}$ is used as a shorthand for $(\tf X_i,\hdots,\tf X_{j})$, $\lceil.\rceil$ denotes the ceiling operator, and $\phi$ denotes null set.  The term $o(n)$ is some function $g(n)$ such that $\underset{n \rightarrow \infty}{\lim} \frac {g(n)}{n}=0$. The dot equality $\doteq$ denotes the equality on the prelog factor, such that for some functions $f(n)$ and $g(n)$, $f(n)\doteq g(n)$ implies $f(n)=g(n)+o(n)$.

\section{System Model and Definitions}
\label{miso-mu-sec-sys}
We consider a multi-user wiretap channel  which consists of  two legitimate receivers and an external eavesdropper as shown in Figure~\ref{model-m}. In this setup, the transmitter is equipped with three transmit antennas and the two receivers and the eavesdropper are equipped with a single antenna each.  The transmitter wants to reliably transmit message $W_{1} \in \mc{W}_{1}=\{1,\hdots,2^{nR_{1}(P)} \}$ to the receiver 1, and message $W_{2} \in \mc{W}_{2}=\{1,\hdots,2^{nR_{2}(P)} \}$ to the receiver 2. In doing so, the transmitter also wishes to conceal both messages $(W_1,W_2)$  from the external eavesdropper. We assume that the external eavesdropper is passive, i.e., it is not allowed to modify the communication.  

We consider a fast fading channel model, and assume that each receiver knows the perfect instantaneous CSI and also the past CSI of the other receiver. The channel input-output relationship at time instant $t$ is given by
\begin{subequations}
 \label{g-chan-mu}
\begin{align}
 y_{1,t} &= {{\tf{h}}}_{t} {{\tf{x}}}_t+n_{1t} \\
 y_{2,t} &= {\acute{\tf{h}}}_{t} {{\tf{x}}}_t+n_{2t} \\
 z_t &= {{\tf{g}}}_t {{\tf{x}}}_t+n_{3t}, \:\: t=1,\hdots,n
\end{align}
\end{subequations}
 where ${{\tf{x}}} \in \mb{C}^{3 \times 1}$ is the channel input vector, ${{\tf{h}}} \in \mc {H} \subseteq \mb{C}^{1 \times 3}$ is the channel vector connecting receiver 1 to the transmitter, ${\acute{\tf{h}}} \in\acute{\mc {H}} \subseteq \mb{C}^{1 \times 3}$ is the channel vector connecting receiver 2 to the transmitter,  and ${{\tf{g}}} \in \mc {G} \subseteq \mb{C}^{1 \times 3}$ is the channel vector connecting the eavesdropper to the transmitter respectively; and $n_{i}$ is assumed to be independent and identically distributed (i.i.d.) white Gaussian noise, with $n_i \sim \mc{CN}(0,1)$ for $i= 1,2,3$.  The channel input is subjected to block power constraints, as $\sum_{t=1}^n \mathbb{E}[\| {{\tf{x}}}_t\|^2] \leq nP$. For ease of exposition, we denote $\tf{S}_t = \left [ \tf{h}_t\:\: {\acute{\tf{h}}}_{t} \:\:  \tf{g}_t \right]^\text{T}$ as the channel state matrix and ${{\tf{S}}}^{t-1} =\{{{\tf{S}}}_1,\hdots,{{\tf{S}}}_{t-1}\}$ denotes the collection of channel state matrices over the past $(t-1)$ symbols respectively. For convenience, we set ${{\tf{S}}}^{0} =\emptyset$. We assume that, at each time instant $t$, the channel state matrix ${{\tf{S}}}_t$ is full rank almost surely. At each time instant $t$, the past states of the channel matrix $\tf{S}^{t-1}$ are known to all terminals. However, the instantaneous states $\tf{h}_t$, ${\acute{\tf{h}}}_{t}$, and $\tf{g}_t$ is known only to the receiver 1,  receiver 2,  and eavesdropper, respectively.  
  
Communication over the wireless channel is particularly sensitive to the quality of  CSIT. Although, there are numerous forms of CSIT, in this work we focus on two of them as follows.
    \begin{enumerate}
    \item \dv{Perfect CSIT}: corresponds to those instances in which the transmitter has perfect knowledge of the instantaneous channel state information. We denote these states by `P'.
    \item \dv{Delayed CSIT}: corresponds to those instances in which at time $t$, the transmitter has perfect knowledge of \ti{only} the past $(t-1)$ channel states. Also, we assume that at time instant $t$ the current channel state is independent of the past $(t-1)$ channel states. We denote these states by `D'.
    \end{enumerate}

 Let $S_1$ denotes the CSIT state of user 1, $S_2$ denotes the CSIT state of user 2 and  $S_3$ denotes the CSIT state of the eavesdropper. Then, based on the availability of the CSIT, the model that we study \eqref{g-chan-mu} belongs to any of the following eight states
 \begin{equation}
 (S_1,S_2,S_3) \in \{PPP, PPD, PDP, PDD, DPP, DPD, DDP, DDD\}.
 \end{equation}
 We denote $\lambda_{S_1S_2S_3}$ be the fraction of time state $S_1S_2S_3$ occurs, such that
 \begin{equation}
 \label{sum-n}
 \sum_{ (S_1,S_2,S_3) \in \{P, D\}^3}\lambda_{S_1S_2S_3} = 1.
 \end{equation}
For simplicity of analysis, we assume that $\lambda_{PDD}=\lambda_{DPD}$, i.e., the fractions of time spent in states $PDD$ and $DPD$ are equal.
 \vspace{.25em}
 
  \begin{definition}\label{def1}
  A code for the  Gaussian $(3,1,1,1)$--multi-receiver wiretap channel with alternating CSIT $(\lambda_{S_1S_2S_3})$ consists of sequence of stochastic encoders at the transmitter,
 \begin{align}
 \label{map-PP}
\{\phi_{1t} \:\: &: \:\: \mc W_{1}{\times}\mc W_{2}{\times}\mc S^{t} \longrightarrow \mc{X}_{1} \times \mc{X}_2 \times \mc{X}_3 \}_{t=1}^{\lceil n\lambda_{PPP}\rceil}\notag\\
\{\phi_{2t} \:\: &: \:\: \mc W_{1}{\times}\mc W_{2}{\times}\mc S^{t-1} {\times}  \mc {H}_{t}  {\times}  \acute{\mc {H}}_{t}  \longrightarrow \mc{X}_{1} \times \mc{X}_2 \times \mc{X}_3\}_{t=1}^{\lceil n\lambda_{PPD}\rceil}\notag\\
\{\phi_{3t} \:\: &: \:\: \mc W_{1}{\times}\mc W_{2}{\times}\mc S^{t-1}{\times}    \mc {H}_{t}  {\times}   \mc {G}_{t} \longrightarrow \mc{X}_{1} \times \mc{X}_2 \times \mc{X}_3\}_{t=1}^{\lceil n\lambda_{PDP} \rceil}\notag\\
\{\phi_{4t} \:\: &: \:\: \mc W_{1} {\times} \mc W_{2}{\times}\mc S^{t-1}{\times}  \mc {H}_{t}    \longrightarrow \mc{X}_{1} \times \mc{X}_2 \times \mc{X}_3\}_{t=1}^{\lceil n\lambda_{PDD}\rceil}\notag\\
\{\phi_{5t} \:\: &: \:\: \mc W_{1}{\times}\mc W_{2}{\times}\mc S^{t-1}{\times}  \acute{\mc {H}}_{t}   {\times} \mc {G}_{t}  \longrightarrow \mc{X}_{1} \times \mc{X}_2 \times \mc{X}_3\}_{t=1}^{\lceil n\lambda_{DPP}\rceil}\notag\\
\{\phi_{6t} \:\: &: \:\: \mc W_{1}{\times}\mc W_{2}{\times}\mc S^{t-1}{\times}   \acute{\mc {H}}_{t}  \longrightarrow \mc{X}_{1} \times \mc{X}_2 \times \mc{X}_3\}_{t=1}^{\lceil n\lambda_{DPD}\rceil}\notag\\
\{\phi_{7t} \:\: &: \:\: \mc W_{1}{\times}\mc W_{2}{\times}\mc S^{t-1}{\times}  \mc {G}_{t}     \longrightarrow \mc{X}_{1} \times \mc{X}_2 \times \mc{X}_3\}_{t=1}^{\lceil n\lambda_{DDP} \rceil}\notag\\
\{\phi_{8t} \:\: &: \:\: \mc W_{1} {\times} \mc W_{2}{\times}\mc S^{t-1}    \longrightarrow \mc{X}_{1} \times \mc{X}_2 \times \mc{X}_3\}_{t=1}^{\lceil n\lambda_{DDD}\rceil}
  \end{align}
  where the messages $W_{1}$ and $W_{2}$ are drawn uniformly over the sets $\mc W_{1}$ and $\mc W_{2}$, respectively; and two decoding functions at the receivers,
  \begin{align}
  \psi_{1} \:\: &: \:\: \mc Y_1^{n}{\times}\mc S^{n-1}{\times} \mc {H}_n \longrightarrow \hat{\mc{W}}_{1}\nonumber\\
  \psi_{2} \:\: &: \:\: \mc Y_2^{n}{\times}\mc S^{n-1}{\times} \acute{\mc {H}}_n \longrightarrow \hat{\mc{W}}_{2}.
  \end{align}
  \end{definition}
  \vspace{.5em}
  \begin{definition}\label{def2}
  A rate pair $(R_{1}(P),R_{2}(P))$ is said to be achievable if there exists a sequence of codes such that,
  \begin{equation}
  \limsup_{n \rightarrow \infty} \text{Pr}\{\hat{W}_{i} \neq W_{i}\}=0, \quad \forall\:\: i \in \{1,2\}.
  \end{equation}
  \end{definition}
 \vspace{.25em}
 
\begin{definition}
 A SDoF pair $(d_{1},d_{2})$ is said to be achievable if there exists a sequence of codes satisfying following,
 \begin{enumerate}
 \item Reliability condition:
 \begin{align}
 \label{rel1}
 & \limsup_{n \rightarrow \infty} \text{Pr}\{\hat{W}_{i} \neq W_{i}\}=0,\quad\quad \forall \:\: i \in \{1,2\},
 \end{align}
 \item Perfect secrecy condition:\footnote{For convenience, with a slight abuse in notations, we replace ${{\tf{S}}}^{n}:= (\tf{S}^{n-1}, \tf{g}_n)$ in~\eqref{sec}.}
 \begin{align}
 \label{sec}
 & \limsup_{n \rightarrow \infty} \frac{I(W_1, W_{2};z^n,{{\tf{S}}}^{n})}{n}=0, 
 \end{align}
 \item and communication rate condition:
 \begin{align}
 \label{com1}
 &\lim_{P \rightarrow \infty} \lim_{n \rightarrow \infty} \frac{\log |\mc W_{i}(n,P)|}{n\log P} \geq d_{i}, \quad \forall\:\: i \in \{1,2\} \end{align}
 \end{enumerate}
 \noindent at receiver 1 and 2, respectively.
 \end{definition}
  \vspace{.5em}
  \begin{definition}\label{definition4}
  We define the SDoF region, $\mc C_{\text{SDoF}}(\lambda_{S_1S_2S_3})$, of the multi-receiver wiretap channel as the set of all achievable non-negative pairs $(d_1, d_2)$.
  \end{definition}

\section{SDoF of the MISO wiretap channel with alternating CSIT}\label{secIII}
In this section, we  consider the special case in which the transmitter wants to send information to the receiver 1, and wishes to conceal it from the eavesdropper.

 \noindent The following theorem characterizes the SDoF of the MISO wiretap channel with alternating CSIT.
 \vspace{.5em}
 \begin{theorem}
 \label{theorem-sdof-wt}
 The SDoF of the (3,1,1)--MISO wiretap channel with alternating CSIT $(\lambda_{S_1S_3})$ is
 \begin{align}
 d_s (\lambda_{S_1S_3})& = 1 - \frac{\lambda_{DD}}{3}.
 \label{eq-sdof-wt}
 \end{align}
 \end{theorem}
 \vspace{.5em}
 \begin{IEEEproof}
 The proof of Theorem 1 appears in Appendix~\ref{proof-sdof-wt}.
 \end{IEEEproof}
 \vspace{.5em}
 \begin{remark}
 The upper bound extends the converse proof of \cite[Theorem 1]{YKPS11} established in the context of SDoF of wiretap channel with delayed CSIT to the case with alternating CSIT. It also uses some elements from the converse proof of \cite{TJSP13} established for the two-user broadcast channel with alternating CSIT by taking imposed security constraints into account. Note that, if delayed channel state information of both receivers is conveyed to the transmitter, i.e., $\lambda_{DD}:=$ 1, the outer bound recovers the SDoF of MISO wiretap channel with delayed CSI \cite[Theorem 1]{YKPS11}.
 We also notice that, the result established in  Theorem~\ref{theorem-sdof-wt} is not restricted to \textit{symmetric} case, i.e., $\lambda_{PD}=\lambda_{DP}$ and holds  in general.
 \end{remark}
 \vspace{.5em}
 
 \begin{remark}
  The achievability proof of Theorem~\ref{theorem-sdof-wt} follows by combining appropriately  fixed CSIT schemes. It is interesting to note that, for a given SDoF, any fixed CSIT scheme can be fully alternated by  other (remaining) fixed schemes. For example, the SDoF of $\frac{2}{3}$ can be achieved by completely using the state $DD$ ($\lambda_{DD}:=1$) or by using any of $PD$, $DP$ or $PP$ state, $\frac{2}{3}$  fraction of communication time.
 \end{remark}

\section{SDoF of Multi-receiver Wiretap channel with  fixed CSIT}\label{fixed}
In this section, we consider the multi-receiver wiretap channel shown in Figure~\ref{model-m} with fixed hybrid CSIT states and establish bounds on SDoF region.
For simplicity of analysis and in accordance with DoF framework, in this work we neglect the effect of additive Gaussian noise at the receivers.

\iffalse
\subsection{2-SDoF using PPP state}
Due to the availability of perfect CSIT from all three nodes, the transmitter can zero-force the information leaked to the unintended receiver and the eavesdropper. Thus, it can be readily shown that one symbol is securely transmitted to each receiver in a single timeslot, yielding 1-SDoF at each receiver.
\fi
\subsection{2-SDoF using PPD state}
In the $PPD$ state, perfect CSIT is available from both legitimate receivers and only past or outdated CSIT is available from the eavesdropper. The following theorem provides the SDoF region of the multi-receiver wiretap channel with the $PPD$ state.
\vspace{.5em}
\begin{theorem}
\label{theorem-sdof-mu-ppd}
The SDoF region of the multi-receiver wiretap channel with the $PPD$ state is given by the set of all non-negative pairs $(d_1,d_2)$ satisfying
\begin{subequations}
\label{ppd-mu}
\begin{align}
\label{eq-ppda}
d_1 &\le 1 \\
\label{eq-ppdb}
d_2 &\le 1\\
\label{eq-ppdc}
d_1+d_2 &\le 2.
\end{align}
\end{subequations}
\end{theorem}
\vspace{.5em}
\begin{IEEEproof}
The converse proof of Theorem~\ref{theorem-sdof-mu-ppd} appears in Appendix~\ref{proof-sdof-mu-ppd-converse}. In what follows, we provide the direct part of the proof that is used to establish Theorem~\ref{theorem-sdof-mu-ppd}. We now show that the SDoF of $(d_1,d_2)=(1,1)$ is achievable. The transmitter wants to send confidential symbols $v$ to the receiver 1 and $w$ to the receiver 2 and wishes to conceal them from the external eavesdropper. In this scheme, the transmitter sends symbols $v$ and $w$ along with the artificial noise $u$ where perfect CSIT from both receivers are utilized in two ways 1) it zero-forces the interference being caused by symbol $w$ intended for the receiver 2 and artificial noise $u$, at the receiver 1 and the interference being caused by symbol $v$ intended for the receiver 1 and artificial noise $u$, at the receiver 2, and in doing so 2) it also secures these two symbols from the external eavesdropper. The transmitter sends
 \begin{eqnarray}
 {\tf{x}}_1 = {\acute{\tf{b}}}_1\left [ v \:\:  \phi \:\: \phi \right ]^\text{T}+{{\tf{b}}}_1\left [ w \:\:  \phi \:\: \phi \right ]^\text{T}+ {{\tf{b}}_{12}}\left [ u \:\:  \phi \:\: \phi \right ]^\text{T}, \:\:\:\:\:\:
 \end{eqnarray}
where ${\acute{\tf{b}}_1}\in \mb{C}^{3\times 1 }$, ${{\tf{b}}_1}\in \mb{C}^{3\times 1 }$, and  ${{\tf{b}}_{12}}\in \mb{C}^{3\times 1 }$  are the precoding vectors chosen such that ${\acute{\tf{h}}}_1{\acute{\tf{b}}_1}=0$, ${{\tf{h}}}_1{{\tf{b}}_1}=0$, and  ${\acute{\tf{h}}}_1{{\tf{b}}_{12}}= {\tf{h}}_1{{\tf{b}}_{12}}=0$. These precoding vectors are known at all nodes. The channel input-output relationship is given by
\begin{subequations}
\begin{align}
y_1 &= {{\tf{h}}}_1{\acute{\tf{b}}_1}v,  \\
y_2 &= {\acute{\tf{h}}}_1{{\tf{b}_1}}w,\\
z &= {{\tf{g}}}_1{\acute{\tf{b}}_1}v + {{\tf{g}}}_1{{\tf{b}}_1}w +{{\tf{g}}}_1 {{\tf{b}}_{12}}u .
\end{align}
\end{subequations} 
At the end of time slot 1, since the receiver 1 knows the CSI $({{\tf{h}}}_1)$ and ${\acute{\tf{b}}_1}$, it decodes the desired symbol $v$ from $y_1$ through channel inversion. The receiver 2 can also perform similar operations to decode the desired symbol $w$. The eavesdropper gets the confidential symbols embedded in with artificial noise and is unable to decode them. The information leaked to the eavesdropper $I(v,w;z|\tf{S})$ can be bounded by
 \begin{align}
 I(v,w;z|\tf{S} )&=     h(z|\tf{S})-h(z|v,w,\tf{S}) \notag\\
 & \le \log(P)-h(u|\tf{S})+o(\log(P))\notag\\
 & \le \log (P) - \log(P)+ o(\log(P))\notag\\
 & = o(\log(P)).
 \end{align}
Thus, 1 symbol is securely send to each receiver over a total of 1 time slot, which yields a SDoF of 1 at each receiver, respectively.
\end{IEEEproof}
\vspace{.5em}

\subsection{3/2-SDoF using PDP state}
In the $PDP$ state, perfect CSIT is available from the receiver 1 and  eavesdropper; and, delayed CSIT is available from the receiver 2.  The following theorem provides the SDoF region of the multi-receiver wiretap channel with the $PDP$ state.
\vspace{.5em}
\begin{theorem}
\label{theorem-sdof-mu-pdp}
The SDoF region of the multi-receiver wiretap channel with the $PDP$ state is given by the set of all non-negative pairs $(d_1,d_2)$ satisfying
\begin{subequations}
\label{pdp}
\begin{align}
\label{eq-pdpa}
d_1 &\le 1 \\
\label{eq-pdpb}
d_1+2d_2 &\le 2.
\end{align}
\end{subequations}

\end{theorem}
\vspace{.5em}
\begin{IEEEproof}
The converse proof of Theorem~\ref{theorem-sdof-mu-pdp} appears in Appendix~\ref{proof-sdof-mu-pdp-converse}. We now provide the coding scheme that shows that the SDoF of $(d_1,d_2)=(1,\frac{1}{2})$ is achievable. In this scheme, the transmitter wants to send two confidential symbols $\tf{v}:=(v_1,v_2)$ to receiver 1 and a confidential symbol $w$ to receiver 2 and wishes to conceal them from the external eavesdropper. The coding scheme comprises of two time slots. In the first time slot the transmitter sends 
 \begin{eqnarray}
 {\tf{x}}_1 = {\tf{b}}_3\left [ v_1 \:\:  v_2 \:\: \phi \right ]^\text{T}+{{\tf{b}}}_{13}\left [ w \:\:  \phi \:\: \phi \right ]^\text{T},
 \end{eqnarray}
where ${{\tf{b}}}_{3}\in \mb{C}^{3\times 1 }$ and ${{\tf{b}}}_{13}\in \mb{C}^{3\times 1 }$ are the precoding vectors chosen such that ${{\tf{g}}}_{1}{{\tf{b}}}_{3}=0$ and ${{\tf{g}}}_{1}{{\tf{b}}}_{13}={{\tf{h}}}_{1}{{\tf{b}}}_{13}=0$. The channel input-output relationship is given by
\begin{subequations}
\begin{align}
y_{1,1} &= {{\tf{h}}}_1{{\tf{b}}}_3\tf{v},  \\
y_{2,1} &= \underbrace{{\acute{\tf{h}}}_1{{\tf{b}}}_3\tf{v}}_{\text{interference}}+{\acute{\tf{h}}}_1{{\tf{b}_{13}}}w,\\
z_1 &= 0 .
\end{align}
\end{subequations} 
At the end of time slot 1, the receiver 1 gets one equation with two unknowns and requires an extra equation to decode the desired symbols. This equation is  being available as interference (side information) at the receiver 2.  Receiver 2 gets the desired symbol $w$ embedded in with some interference $({\acute{\tf{h}}}_1{{\tf{b}}}_3\tf{v})$.
Conveying this interference \ti{securely} to both legitimate receivers will be useful in two ways, 1) it provides the extra equation to the receiver 1 to decode the desired symbols $\tf{v}$, and 2) also helps the receiver 2 to remove the interference from $y_{2,1}$ to decode $w$. Due to the availability of delayed  CSI from the receiver 2 $(\acute{\tf{h}})$ and since the transmitter knows $\tf{v}$, it can readily construct ${\acute{\tf{h}}}_1{{\tf{b}}}_3\tf{v}$ and sends 
 \begin{eqnarray}
 {\tf{x}}_2 = {\tf{b}}\left [ 
 {\acute{\tf{h}}}_1{{\tf{b}}}_3\tf{v} \:\:  \phi \:\: \phi \right ]^\text{T}
 \end{eqnarray}
where ${{\tf{b}}}\in \mb{C}^{3\times 1 }$ is the precoding vector chosen such that $\tf{g}_2\tf{b}= 0$.
The channel input-output relationship is given by
\begin{subequations}
\begin{align}
y_{1,2} &= {{\tf{h}}}_2{{\tf{b}}}{\acute{\tf{h}}}_1{{\tf{b}}}_3\tf{v},  \\
y_{2,2} &=\acute{{\tf{h}}}_2{{\tf{b}}}{\acute{\tf{h}}}_1{{\tf{b}}}_3\tf{v},\\
z_2 &= 0.
\end{align}
\end{subequations} 
At the end of time slot 2, since the receiver 1 knows the CSI, it decodes $(v_1,v_2)$ from $(y_{1,1},y_{1,2})$ through channel inversion. Similarly, since the receiver 2 knows the CSI and $y_{2,2}$, it  subtracts out the contribution of ${\acute{\tf{h}}}_1{{\tf{b}}}_3\tf{v}$ from $y_{2,1}$ to decode $w$. The eavesdropper is unable to get any information from the two time slots and thus the information leaked to the eavesdropper $I(\tf{v},w;z_1,z_2|\tf{S}^n)=0$.

It can be readily seen from the above analysis that 2 symbols are securely send  to the receiver 1 over a total of 2 time slots, which yields a SDoF of 1 at the receiver 1. Similarly, 1 symbol is send to the receiver 2 over a total of 2 time slots, which yields a SDoF of $\frac{1}{2}$ at the receiver 2.
\end{IEEEproof}
\vspace{.5em}

\subsection{4/3-SDoF using DDP state}
In this state, perfect CSIT is available from the eavesdropper and only past or outdated CSIT is available from both legitimate receivers. The following theorem provides the SDoF region of the multi-receiver wiretap channel with the $DDP$ state.
\vspace{.5em}
\begin{theorem}
\label{theorem-sdof-mu-ddp}
The SDoF region of the multi-receiver wiretap channel with the $DDP$ state is given by the set of all non-negative pairs $(d_1,d_2)$ satisfying
\begin{subequations}
\label{ddp}
\begin{align}
\label{eq-ddpa}
d_1+2d_2 &\le 2\\
\label{eq-ddpb}
2d_1+d_2 &\le 2.
\end{align}
\end{subequations}
\end{theorem}
\vspace{.5em}
\begin{IEEEproof}
The converse proof of Theorem~\ref{theorem-sdof-mu-ddp} follows along very similar lines as in~\eqref{eq-pdpb} and is omitted for brevity. We now provide the direct part of the proof that is used to establish Theorem~\ref{theorem-sdof-mu-ddp} and  show that the SDoF of $(d_1,d_2)=(\frac{2}{3},\frac{2}{3})$ is achievable. In this scheme, the transmitter wants to send two confidential symbols $\tf{v}:= (v_1,v_2)$ to receiver 1 and  two confidential symbols $\tf{w}:=(w_1,w_2)$ to receiver 2 and wishes to conceal them from the eavesdropper. The coding scheme comprises of three time slots. In the first time slot the transmitter sends 
 \begin{eqnarray}
 {\tf{x}}_1 = {\tf{b}}_1\left [ v_1 \:\:  v_2 \:\: \phi \right ]^\text{T},
 \end{eqnarray}
where ${{\tf{b}}}_{1}\in\mb{C}^{3\times 1 }$ is the precoding vector chosen such that ${{\tf{g}}}_{1}{{\tf{b}}}_{1}=0$.   The channel input-output relationship is given by
\begin{subequations}
\begin{align}
y_{1,1} &= {{\tf{h}}}_1{{\tf{b}}}_1\tf{v},  \\
y_{2,1} &=  \underbrace{{\acute{\tf{h}}}_1{{\tf{b}}}_1\tf{v}}_{\text{interference}},\\
z_1 &= 0 .
\end{align}
\end{subequations} 
At the end of time slot 1, both receivers convey the past CSI to the transmitter. At the end of time slot 1, the receiver 1 gets one equation with two unknowns and requires an extra equation to decode the desired symbols. This equation is  being available as interference (side information) at the receiver 2.  If this interference can be conveyed to the receiver 1, it suffices to decode $\tf{v}$. 

In the second time slot the transmitter sends fresh information to the receiver 2 as
 \begin{eqnarray}
 {\tf{x}}_2 = {\tf{b}}_2\left [ w_1 \:\:  w_2 \:\: \phi \right ]^\text{T},
 \end{eqnarray}
where ${{\tf{b}}}_{2}\in\mb{C}^{3\times 1 }$ is the precoding vector chosen such that ${{\tf{g}}}_{2}{{\tf{b}}}_{2}=0$.  The channel input-output relationship is given by
\begin{subequations}
\begin{align}
y_{1,2} &= \underbrace{{{\tf{h}}}_2{{\tf{b}}}_2\tf{w}}_{\text{interference}},  \\
y_{2,2} &=  {\acute{\tf{h}}}_2{{\tf{b}}}_2\tf{w},\\
z_2 &= 0 .
\end{align}
\end{subequations} 
At the end of time slot 2, both receivers convey the past CSI to the transmitter. At the end of time slot 2, the receiver 2 gets one equation with two unknowns and requires an extra equation to decode the desired symbols. This equation is  being available as interference (side information) at the receiver 1.  Conveying this interference to the receiver 2 suffices to decode $\tf{w}$. 

Due to the availability of past CSI, the transmitter is able to construct the side information required by the receiver 1 available at receiver 2 in time slot 1, $y_{2,1}$, and side information required by receiver 2 available at the receiver 1 in time slot 2, $y_{1,2}$. In the third time slot, the transmitter sends
 \begin{eqnarray}
 {\tf{x}}_3 = {\tf{b}}_3\left [  {{\tf{h}}}_2{{\tf{b}}}_2\tf{w}+{\acute{\tf{h}}}_1{{\tf{b}}}_1\tf{v} \:\:  \phi \:\: \phi \right ]^\text{T}
 \end{eqnarray}
where ${{\tf{b}}}_3\in\mb{C}^{3\times 1 }$ is the precoding vector chosen such that $\tf{g}_3\tf{b}_3= 0$.  The channel input-output relationship is given by
\begin{subequations}
\begin{align}
y_{1,3} &= \tf{h}_3\tf{b}_3({{\tf{h}}}_2{{\tf{b}}}_2\tf{w}+{\acute{\tf{h}}}_1{{\tf{b}}}_1\tf{v}),  \\
y_{2,3} &=\acute{\tf{h}}_3\tf{b}_3({{\tf{h}}}_2{{\tf{b}}}_2\tf{w}+{\acute{\tf{h}}}_1{{\tf{b}}}_1\tf{v}),\\
z_3 &= 0.
\end{align}
\end{subequations} 
At the end of time slot 3, since the receiver 1 knows the CSI and $y_{1,2}$, it first subtracts out the contribution of $y_{1,2}$ from $y_{1,3}$; and decodes $(v_1,v_2)$ from $(y_{1,1},y_{1,3})$ through channel inversion. Similarly, since the receiver 2 knows the CSI and $y_{2,1}$, it subtracts out the contribution of $y_{2,1}$ from $y_{2,3}$ to decode $(w_1,w_2)$. 

From the above analysis, it can be easily seen that 2 symbols are securely send to the $i$-th receiver over a total of 3 time slots, which yields a SDoF of $\frac{2}{3}$ at each receiver, $i=1,2$, respectively.
\end{IEEEproof}
\vspace{.5em}

\subsection{1-SDoF using PDD state}
\label{pd1}
In this state perfect CSIT is available from the receiver 1 and delayed or past CSIT is available from the receiver 2 and the eavesdropper. For this state, we now show that the sum SDoF of 1 is achievable. The transmitter sends a confidential symbol $v$ intended for the receiver 1 along with artificial noise $u$
as
 \begin{eqnarray}
 {\tf{x}}_1 = \left [ v \:\: \phi \:\: \phi \right ]^\text{T}+{{\tf{b}}}_1\left [ u \:\:  \phi \:\: \phi \right ]^\text{T},
 \end{eqnarray}
where the precoding vector ${{\tf{b}}_1}\in\mb{C}^{3\times 1 }$ is chosen such that  ${{\tf{h}}}_1{{\tf{b}}_1}=0$. Thus, receiver 1 can easily decode the desired symbol. The eavesdropper gets the confidential symbol embedded in with artificial noise and thus is unable to decode it.  Thus, 1 symbol is securely send to the receiver 1 over a total of 1 time slot,  yielding  the SDoF pair $(d_1,d_2)=(1,0)$.

\section{SDoF of Multi-Receiver Wiretap channel with alternating CSIT}
\label{alt}
We now turn our attention to the multi-receiver wiretap channel, in which the transmitter is allowed to alternate between two states, i.e., $PDD$ and $DPD$, equal fractions of the communication time.

\subsection{Outer Bound}
The following theorem provides an outer bound on the SDoF region of the multi-receiver wiretap  channel with alternating CSIT.
\vspace{.5em}
\begin{theorem}
\label{theorem-sdof-mu-main-outer-bound}
An outer bound on the SDoF region $\mc C_\text{SDoF}(\lambda_{S_1S_2S_3})$ of the multi-receiver wiretap channel with alternating CSIT is given by the set of all non-negative pairs $(d_1,d_2)$ satisfying
\begin{subequations}
\label{theorem-sdof-mu-outer-bound-eqautions}
\begin{align}
\label{miso-mu-outer-1}
16d_1 + 4d_2  &\leq 17\\
\label{miso-mu-outer-2}
4d_1 + 16d_2  &\leq 17.
\end{align}
\end{subequations}
\end{theorem}
\vspace{.5em}
\begin{IEEEproof}
The proof of Theorem~\ref{theorem-sdof-mu-main-outer-bound} appears in Appendix~\ref{proof-sdof-mu-outer-bound}.
\end{IEEEproof}
\vspace{.5em}
  
\subsection{Inner Bound}
\noindent Next, we establish an inner bound on the multi-receiver wiretap channel with alternating CSIT.
\vspace{.5em}
\begin{theorem}
\label{theorem-sdof-mu-achievability}
An inner bound on the SDoF region $\mc C_\text{SDoF}(\lambda_{S_1S_2S_3})$ of the multi-receiver MISO wiretap channel with alternating CSIT is given by the set of all non-negative pairs $(d_1,d_2)$ satisfying
\begin{subequations}
\label{theorem-sdof-mu-bc-achievability-eqautions}
\begin{align}
\label{miso-mu-acheivbaility-1}
15d_1+14d_2 &\leq 15\\
\label{miso-mu-acheivbaility-2}
14d_1+15d_2&\leq 15.
\end{align}
\end{subequations}
\end{theorem}
\vspace{.5em}
\begin{IEEEproof}
The region in~\eqref{theorem-sdof-mu-bc-achievability-eqautions} is characterized by the corner points $(1,0)$, $(0,1)$ and the point $(15/29,15/29)$ obtained by the intersection of line equations in~\eqref{theorem-sdof-mu-bc-achievability-eqautions}. The achievability of the two corner points $(1,0)$  and  $(0,1)$ follow by the coding scheme developed in 
Theorem~\ref{theorem-sdof-wt}, where the transmitter is interested to send confidential message to the receiver 1 being eavesdropped by the  eavesdropper. The achievability of the point  $(15/29,15/29)$  is provided in subsection~\ref{proof-mu}.
\end{IEEEproof}
\vspace{.5em}

\begin{figure}
 \psfragscanon
 \centering
 \psfrag{data1}[c][c][.55]{\hspace{25em}Achievable SDoF   with fixed hybrid  CSIT ($PDD$ or $DPD$ state)}
 \psfrag{data3}[c][c][.55]{\hspace{22em}\:\:Outer bound on SDoF  with $PDD/DPD$ state Theorem~\ref{theorem-sdof-mu-main-outer-bound}}
 \psfrag{data2}[c][c][.55]{\hspace{22em}\:\:Inner bound on SDoF  with $PDD/DPD$ state Theorem~\ref{theorem-sdof-mu-achievability}} 
 \psfrag{A}[c][c][.8]{\hspace{-3em}$(\frac{17}{20},\frac{17}{20})$}
 \psfrag{B}[c][c][.8]{\hspace{-3em}$(\frac{15}{29},\frac{15}{29})$}
 \psfrag{m}[c][c]{\vspace{1em}$d_1$}
 \psfrag{n}[c][c]{\vspace{1em}$d_2$}
 \includegraphics[width=1\linewidth]{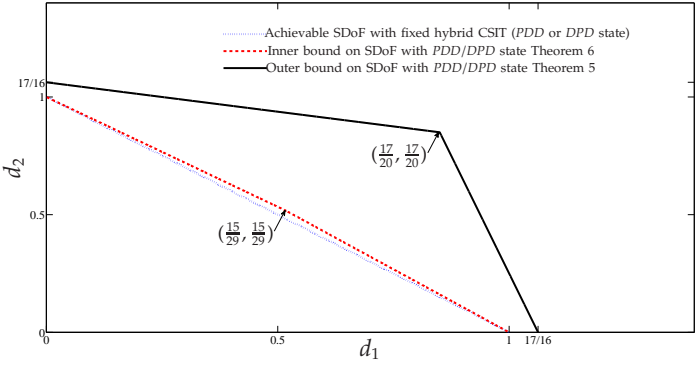}
 \caption{SDoF region of multi-receiver wiretap channel with alternating CSIT.}
 \psfragscanoff
 \label{MISO-mu-comp}
 \end{figure}
Figure~\ref{MISO-mu-comp} shows the outer and inner bounds on SDoF with alternating CSIT in~\eqref {theorem-sdof-mu-outer-bound-eqautions} and~\eqref{theorem-sdof-mu-bc-achievability-eqautions}, respectively. For comparison reasons, we also plot the SDoF region obtained by fixed state $PDD$. It can be easily seen from Figure~\ref{MISO-mu-comp} that, by synergistically using $PDD$ and $DPD$ states the inner bound in~\eqref{theorem-sdof-mu-bc-achievability-eqautions} provides a sum rate
 \begin{eqnarray}
 \text{SDoF}_{\text{sum}}=  \underbrace{\frac{30}{29}}_{\text{PDD/DPD}} \ge  \underbrace{1}_{\text{PDD}}
 \end{eqnarray}
which is clearly larger than the sum rate with fixed CSIT state.

\subsection{S$_1^{30/29}$ --- Coding scheme using $PDD$ and $DPD$ states}
\label{proof-mu}
We now provide some coding schemes that provide the main ingredients to establish the inner bound in Theorem~\ref{theorem-sdof-mu-achievability}. The following schemes achieve $30/29$ SDoF.
 \begin{enumerate}
 \item $S_1^{30/29}$ -- using $PDD$, $DPD$ states for $(\frac{22}{29},\frac{7}{29})$ fractions of time, $(d_1,d_2)= (\frac{15}{29},\frac{15}{29})$ SDoF is achievable.
 \item $S_2^{30/29}$ -- using $PDD$, $DPD$ states for $(\frac{7}{29},\frac{22}{29})$ fractions of time, $(d_1,d_2)= (\frac{15}{29},\frac{15}{29})$ SDoF is achievable.
 \end{enumerate}
 \vspace{.5em}
The achievability of the corner point  $(15/29,15/29)$ in Theorem~\ref{theorem-sdof-mu-achievability} follows by using $S_1^{30/29}$ and $S_2^{30/29}$ schemes equal fractions of communication time.

\vspace{.5em}
\subsubsection{S$_1^{30/29}$ --- Coding scheme using $PDD$ and $DPD$ states $(\frac{22}{29},\frac{7}{29})$ fractions of time}
We now show that by using $PDD$ and $DPD$ states for $(\frac{22}{29},\frac{7}{29})$ fractions of time, $(d_1,d_2)= (\frac{15}{29},\frac{15}{29})$ SDoF is achievable. In this scheme, the transmitter wants to transmit three symbols $(v_1,v_2,v_3)$ to the receiver 1 and three symbols ($w_1,w_2,w_3$) to the receiver 2 and wishes to conceal them from the eavesdropper. The communication takes place in two phases, i.e., data dissemination phase and transmission of common information.

 \vspace{.25em}
\ti{A) Data dissemination phase:}
In this phase the transmitter sends fresh information to both receivers. In the first time slot, the transmitter chooses $PDD$ state and injects artificial noise  $\tf{u}:=[u_1,u_2,u_3]^T$, from all antennas.
At the end of time slot 1, the channel input-output relationship is given by
\begin{subequations}
\begin{align}
y_{1,1} &=  \tf{h}_1  {\tf{u}}, \\
y_{2,1} &=  \acute{\tf{h}}_1 \tf{u} , \\
z_{1} &=  \tf{g}_1  \tf{u} .
\end{align}
\end{subequations}
At the end of time slot 1, the receiver 2 and eavesdropper feed back the delayed CSI to the transmitter. 

In the second time slot, the transmitter remains in $PDD$ state and sends fresh information $\tf{v
}:=[v_1, v_2,v_3]^T$ to the receiver 1 along with a linear combination of channel output $y_{1,1}$ at the receiver 1. The transmitter can easily learn $y_{1,1}$, since it already knows the perfect CSI $(\tf{h}_1)$ and $\tf{u}$. During this phase, the transmitter sends
\begin{eqnarray}
{\tf{x}}_2 = \left [ v_1 \:\:  v_2 \:\: v_3 \right ]^\text{T}+ \left [ y_{1,1} \:\:  \phi \:\: \phi \right ]^\text{T}.
\end{eqnarray}
At the end of time slot 2, the channel input-output relationship is given by
\begin{subequations}
\begin{align}
y_{1,2} &=   \tf{h}_2 \tf{v} +h_{21}y_{1,1}, \\
y_{2,2} &= \underbrace{\acute{\tf{h}}_2 \tf{v}+ \acute{{h}}_{21}y_{1,1}}_{\text{side information}}, \\
z_{2}   &= \underbrace{\tf{g}_2 \tf{v}+g_{21}y_{1,1}}_{\text{side information}}.
\end{align}
\end{subequations}
At the end of time slot 2, the receiver 2 and eavesdropper feed back the delayed CSI to the transmitter. At the end of time slot 2, the receiver 1 can subtracts out the contribution of $y_{1,1}$ to get one equation with 3 confidential symbols and requires two extra equations to successfully decode the intended variables. This side information is available at the receiver 2 and eavesdropper, and will be conveyed in phase 2. 

In the third time slot, the transmitter remains in $PDD$ state and sends fresh information $\tf{w
}:=[w_1, w_2,w_3]^T$ to the receiver 2 along with a linear combination of channel output $y_{2,1}$ at the receiver 2 at the end of first time slot. The transmitter can easily re-construct $y_{2,1}$, since it  knows the past CSI $(\acute{\tf{h}}_1)$ and $\tf{u}$. During this phase, the transmitter sends
\begin{eqnarray}
{\tf{x}}_3 = \left [ w_1 \:\:  w_2 \:\: w_3 \right ]^\text{T}+ \left [ y_{2,1} \:\:  \phi \:\: \phi \right ]^\text{T}.
\end{eqnarray}
The channel input-output relationship is given by
\begin{subequations}
\begin{align}
y_{1,3} &=   \underbrace{ \tf{h}_3 \tf{w}+  {{h}}_{31}y_{2,1}}_{\text{side information}}, \\
y_{2,3} &=  \acute{\tf{h}}_3 \tf{w}+ \acute{{h}}_{31}y_{2,1}, \\
z_{3} &= \underbrace{\tf{g}_3 \tf{w}+g_{31}y_{2,1}}_{\text{side information}}.
\end{align}
\end{subequations}
At the end of time slot 3, the receiver 2 and eavesdropper feed back the delayed CSI to the transmitter. At the end of time slot 3, the receiver 2  subtracts out the contribution of $y_{2,1}$ to get one equation with 3 confidential symbols and requires two extra equations to successfully decode the intended variables. This side information is available at the receiver 1 and eavesdropper, respectively. 

Recall that, at the end of three time slots, the receiver 1 requires side information available at the receiver 2 $(y_{2,2})$ and eavesdropper $(z_{2})$ and the receiver 2 requires side information available at the receiver 1 $(y_{1,3})$ and eavesdropper $(z_3)$. Due to the availability of non-causal and strictly causal CSIT, the transmitter can learn these side informations and the next step is how to convey them securely. The information leaked to eavesdropper after 3 time slots is bounded by
 \begin{eqnarray}
  && \hspace{-5em}I(\tf{v},\tf{w};z_1,z_2,z_3|\tf{S}^n)  \notag\\&&\hspace{-4em} \le  I(\tf{g}_2\tf{v},\tf{g}_3\tf{w};z_1,z_2,z_3|\tf{S}^n) \notag\\
  && \hspace{-4em} =  I(\tf{g}_2\tf{v},\tf{g}_3\tf{w},\tf{u};z_1,z_2,z_3|\tf{S}^n)\notag\\&&\hspace{-4em} -  I(\tf{u};z_1,z_2,z_3|\tf{g}_2\tf{v},\tf{g}_3\tf{w},\tf{S}^n)\notag\\
  && \hspace{-4em}  \le  I(\tf{g}_2\tf{v}+g_{21}\tf{h}_1\tf{u},\tf{g}_3\tf{w}+g_{31}\tf{g}_1\tf{u}, \tf{g}_1\tf{u};z_1,z_2,z_3|\tf{S}^n)\notag\\&&\hspace{-4em} -  I(\tf{u};z_1,z_2,z_3|\tf{g}_2\tf{v},\tf{g}_3\tf{w},\tf{S}^n)\notag\\
 && \hspace{-4em} \overset{(a)}{=} 3\log (P) - 3\log(P)+o(\log(P))\notag\\
  && \hspace{-4em}  = o(\log(P))
 \end{eqnarray}
 where $(a)$ follows from~\cite[Lemma 2]{YKPS11}. 
 
\noindent The side information available at the eavesdropper can be conveyed in the spirit of alternating CSIT scheme developed in~\cite[Theorem 1]{TJSP13} where alternation between $PD$ and $DP$ states equal fractions of communication time yields an optimal $\text{DoF}_{\text{PD/DP}} = 5/3$. Thus, the side information required by the receiver 1 $(z_2)$ and the receiver 2 $(z_3)$ can be conveyed to both receivers over a total of $\frac{2}{\text{DoF}_{\text{PD/DP}}} = 6/5$ time slots. After conveying these side informations to respective receivers, the receiver 1 requires $y_{2,2}$ which is available at the receiver 2 at the end of time slot 2 and the receiver 2 requires $y_{1,3}$ which is available at the receiver 1 at the end of time slot 3 to successfully decode the desired symbols. Note that, one can not merely multicast these side information similar to $\text{DoF}_{\text{PD/DP}}$~\cite[Theorem 1]{TJSP13}, since it will leak extra information to the eavesdropper. Next, we define a common message $W_{12} := y_{2,2}+y_{1,3}$. Conveying $W_{12}$ to both receivers securely will suffice to decode their respective symbols. The resulting SDoF at each receiver can be concisely written as 
\begin{eqnarray}
\label{sum-common}
d_i = \frac{3}{3+\frac{2}{\text{DoF}_{\text{PD/DP}}}+\frac{1}{\text{SDoF}_{\text{common}}}}, \quad i=1,2
\end{eqnarray}
where $\text{SDoF}_{\text{common}}$ denotes the SDoF of the common message $W_{12}$.

 \vspace{.5em}
\ti{B)  Multicasting common information with alternating CSIT:}
We now provide the description of the coding scheme which is used to send two common symbols  $v_{12}$ and $w_{12}$ over a total of $\frac{16}{5}$ time slots to both receivers with alternating CSIT, securely. In the first time slot,  the transmitter chooses  $PDD$ state and transmits the confidential symbol $v_{12}$ embedded in with artificial noise $q_1$ as
\begin{eqnarray}
{\tf{x}}_1 = \left [ v_{12} \:\:  \phi \:\: \phi \right ]^\text{T}+ \tf{b}_1\left [ q_1 \:\:  \phi \:\: \phi \right ]^\text{T}, 
\end{eqnarray}
where $\tf{b}_1 \in \mb{C}^{3\times 1 }$ is the precoding vector chosen such that ${\tf{h}}_1 {\tf{b}_1}$ = 0. At the end of timeslot 1, the channel input-output relationship is given by
\begin{subequations}
\begin{align}
y_{1,1} &= {h}_{11} v_{12}, \\
y_{2,1} &= \acute{{h}}_{11}v_{12} +\acute{\tf{h}}_1\tf{b}_1 q_1, \\
z_1 &= \underbrace{g_{11}v_{12} + \tf{g}_1\tf{b}_1 q_1}_{\text{side information}}.
\end{align}
\end{subequations}
At the end of time slot 1,  the receiver 1 can readily decode symbol $v_{12}$ through channel inversion. Receiver 2 gets the confidential symbol embedded in with artificial noise $q_1$ and requires one extra equation to decode $v_{12}$. This side information is available at the eavesdropper.

In the second time slot,  the transmitter switches to $DPD$ state and transmits the confidential symbol $w_{12}$ embedded in with artificial noise $q_2$ as
\begin{eqnarray}
{\tf{x}}_2 = \left [ w_{12} \:\:  \phi \:\: \phi \right ]^\text{T}+ \tf{b}_2\left [ q_2 \:\:  \phi \:\: \phi \right ]^\text{T},
\end{eqnarray}
where $\tf{b}_2 \in \mb{C}^{3\times 1 }$ is the precoding vector chosen such that ${\acute{\tf{h}}}_2 {\tf{b}_2}$ = 0. At the end of timeslot 2, the channel input-output relationship is given by
\begin{subequations}
\begin{align}
y_{1,2} &= {h}_{21} w_{12}+ {\tf{h}}_2\tf{b}_2 q_2,  \\
y_{2,2} &= \acute{{h}}_{21}w_{12},\\
z_2 &= \underbrace{g_{21}w_{12} + \tf{g}_2\tf{b}_2 q_2}_{\text{side information}}.
\end{align}
\end{subequations}
At the end of time slot 2,  the receiver 2 can readily decode symbol $w_{12}$ through channel inversion. Receiver 1 gets the confidential symbol embedded in with artificial noise $q_2$ and requires one extra equation to decode $w_{12}$. 

At the end of two time slots, both receivers require one extra equation to decode their respective messages being available at the eavesdropper. By using $\text{DoF}_{\text{PD/DP}}$ scheme $z_1$ is send to the receiver 2 and $z_2$ is send to the receiver 1 over a total of $\frac{2}{\text{DoF}_{\text{PD/DP}}} = 6/5$ time slots. Thus, 2 symbols are securely send to both receivers over a total of $2+6/5$ time slots which yields a SDoF of 
\begin{eqnarray}
\label{common}
\text{SDoF}_{\text{common}} &=  \frac{2}{2+\frac{6}{5}}\notag\\
&= \frac{5}{8}.
\end{eqnarray}

\noindent Finally replacing~\eqref{common} in~\eqref{sum-common} yields the SDoF of 
\begin{eqnarray}
\label{sum-common}
d_i &= \frac{3}{3+\frac{2}{5/3}+\frac{1}{5/8}}
&= \frac{15}{29}
\end{eqnarray}
at each receiver securely.

\subsubsection{S$_2^{30/29}$ --- Coding scheme using $PDD$ and $DPD$ states $(\frac{7}{29},\frac{22}{29})$ fractions of time}
The coding scheme in this case follows along similar lines as the scheme illustrated above by reversing the roles of receiver 1 and receiver 2, respectively.

 \section{SDoF of the MISO broadcast channel with alternating CSIT}
 \label{secIV}
 In this section, we consider the two-user multiple-input single-output (MISO) broadcast channel, as shown in Figure~\ref{model}. In this setting, the transmitter is equipped with two transmit antennas and the two receivers are equipped with a single antenna each. The transmitter wants to reliably transmit message $W_{1} \in \mc{W}_{1}=\{1,\hdots,2^{nR_{1}(P)} \}$ to receiver 1, and message $W_{2} \in \mc{W}_{2}=\{1,\hdots,2^{nR_{2}(P)} \}$ to receiver 2. In doing so, the transmitter also wishes to conceal the message $W_{1}$ that is intended to the receiver 1 from the receiver 2; and the message $W_{2}$ that is intended to the receiver 2 from the receiver 1. Thus, in the considered system configuration, the receiver 2 acts as an eavesdropper on the MISO channel to receiver 1; and receiver 1 acts an eavesdropper on the MISO channel to receiver 2.  
 
 The channel input-output relationship at time instant $t$ is given by
  \begin{align}
  \label{g-chan}
  y_{1,t} &= {{\tf{h}}}_{1t} {{\tf{x}}}_t+n_{1t} \notag\\
  z_t &= {{\tf{g}}}_t {{\tf{x}}}_t+n_{3t}, \:\: t=1,\hdots,n
  \end{align}
  where ${{\tf{x}}} \in \mb{C}^{2 \times 1}$ is the channel input vector, ${{\tf{h}_1}} \in \mc {H}_1 \subseteq \mb{C}^{1 \times 2}$ is the channel vector connecting receiver 1 to the transmitter and ${{\tf{g}}} \in \mc {G} \subseteq \mb{C}^{1 \times 2}$ is the channel vector connecting receiver 2 to the transmitter respectively; and $n_{i}$ is assumed to be independent and identically distributed (i.i.d.) white Gaussian noise, with $n_i \sim \mc{CN}(0,1)$ for $i= 1,3$. The channel input is subjected to block power constraints, as  $\sum_{t=1}^n \mathbb{E}[\| {{\tf{x}}}_t\|^2] \leq nP$. 
  
 Let $S_1$ denotes the CSIT state of user 1 and $S_2$ denotes the CSIT state of user 2. Then, based on the availability of the CSIT, the model \eqref{g-chan} belongs to any of the four states $(S_1,S_2) \in \{P,D\}^2$.  We denote $\lambda_{S_1S_2}$ be the fraction of time state $S_1S_2$ occurs, such that
  \begin{equation}
  \label{sum}
  \sum_{(S_1,S_2) \in \{P,D\}^2}\lambda_{S_1S_2} = 1.
  \end{equation}
  Also, due to the symmetry of problem as reasoned in~\cite{TJSP13}, in this model we assume that $\lambda_{PD}=\lambda_{DP}$, i.e., the fractions of time spent in state $PD$ and $DP$ are equal.
  \vspace{.5em}
  
  \begin{definition}\label{def3}
  A SDoF pair $(d_{1},d_{2})$ is said to be achievable if there exists a sequence of codes satisfying following,
  \begin{enumerate}
  \item Reliability condition~\eqref{rel1}
   
  \item Perfect secrecy condition:\footnote{With a slight abuse in notations, we replace ${{\tf{S}}}^{n}:= (\tf{S}^{n-1}, \tf{h}_n)$,  ${{\tf{S}}}^{n}:= (\tf{S}^{n-1}, \tf{g}_n)$ in~\eqref{sec-constraint-1} and~\eqref{sec-constraint-2}, respectively.}
\begin{subequations}
\begin{align}
  \label{sec-constraint-1} 
  & \limsup_{n \rightarrow \infty} \frac{I(W_{2};y_1^n,{{\tf{S}}}^{n})}{n}=0,\\
   \label{sec-constraint-2} 
  & \limsup_{n \rightarrow \infty} \frac{I(W_{1};z^n,{{\tf{S}}}^{n})}{n}=0,
  \end{align}
\end{subequations}
  \item and Communication rate condition~\eqref{com1} at both receivers.
  \end{enumerate}
  
  \end{definition}

The rest of the system description is similar to the model defined in Section~\ref{miso-mu-sec-sys}, and, so we omit the details for
brevity.

 \subsection{Outer Bound}
  The following theorem provides an outer bound on the SDoF region of the MISO broadcast channel with alternating CSIT.
 \vspace{.5em}
 \begin{theorem}
 \label{theorem-sdof-miso-bc-outer-bound}
 An outer bound on the SDoF region $\mc C_\text{SDoF}(\lambda_{S_1S_2})$ of the two user (2,1,1)--MISO broadcast channel with alternating CSIT is given by the set of all non-negative pairs $(d_1,d_2)$ satisfying
 \begin{subequations}
 \label{theorem-sdof-miso-bc-outer-bound-eqautions}
 \begin{align}
 \label{miso-bc-outer-a}
 d_1 &\le d_s \\
 \label{miso-bc-outer-b}
 d_2 &\le d_s \\
 \label{miso-bc-outer-1}
 3d_1 + d_2  &\leq 2+ 2 \lambda_{PP}+2\lambda_{PD} \\
 \label{miso-bc-outer-2}
  d_1 + 3 d_2  &\leq 2+ 2 \lambda_{PP}+2\lambda_{PD} .
 \end{align}
 \end{subequations}
 \end{theorem}
 \vspace{.5em}
 \begin{IEEEproof}
 The outer bound follows by generalizing the converse that we have established in  Theorem~\ref{theorem-sdof-wt} in the context of the MISO wiretap channel with alternating CSIT to the broadcast setting. The proof of the upper bounds~\eqref{miso-bc-outer-a} and~\eqref{miso-bc-outer-b}  in Theorem~\ref{theorem-sdof-miso-bc-outer-bound} follows along the lines of  the one established in Theorem~\ref{theorem-sdof-wt}. The proof of~\eqref{miso-bc-outer-1} and~\eqref{miso-bc-outer-2} is provided in~\cite{AZS-15}.
 \end{IEEEproof}
 \vspace{.5em}

 \subsection{Inner Bound}
 \noindent We now establish an inner bound on the MISO broadcast channel with alternating CSIT $(\lambda_{S_1S_2})$. For convenience, we first define the following quantity
 \begin{equation}
 d_s^{\text{low}}= d_s-\frac{6\lambda_{PD}}{11}.
 \end{equation}
 \noindent The following theorem provides an inner bound on the SDoF region of the MISO broadcast channel with alternating CSIT.
 \vspace{.5em}
 \begin{theorem}
 \label{theorem-sdof-miso-bc-achievability}
 An inner bound on the SDoF region $\mc C_\text{SDoF}(\lambda_{S_1S_2})$ of the two user (2,1,1)--MISO broadcast channel with alternating CSIT is given by the set of all non-negative pairs $(d_1,d_2)$ satisfying
 \begin{subequations}
 \label{theorem-sdof-miso-bc-achievability-eqautions}
 \begin{align}
 d_1 &\le d_s \\
 d_2 &\le d_s \\
 \label{miso-bc-acheivbaility-1}
 \frac{d_1}{d_s^{\text{low}}}+\frac{d_2}{2} &\leq 1+ \frac{\lambda_{PP}+\lambda_{PD}}{2}\\
 \label{miso-bc-acheivbaility-2}
 \frac{d_1}{2}+\frac{d_2}{d_s^{\text{low}}} &\leq 1 + \frac{\lambda_{PP}+\lambda_{PD}}{2}.
 \end{align}
 \end{subequations}
 \end{theorem}
 \vspace{.5em}
 \begin{IEEEproof}
 The proof of Theorem~\ref{theorem-sdof-miso-bc-achievability} appears in~\cite{AZS-15}.
 \end{IEEEproof}
 \vspace{.5em}
 \begin{remark}
 The region established in Theorem~\ref{theorem-sdof-miso-bc-achievability} reduces to the DoF region of the MISO broadcast channel with alternating CSIT
 and no security constraints in~\cite[Theorem 1]{TJSP13} by setting $d_s=d_s^{\text{low}}:= 1$ in~\eqref{theorem-sdof-miso-bc-achievability-eqautions}.  
 For the special case in which instantaneous or delayed CSI is conveyed by \ti{both} receivers, i.e., $\lambda_{PP}:=1$ or $\lambda_{DD}:=1$, respectively, the outer and inner bounds coincide and SDoF region is established.  
 \end{remark}
 \begin{figure}
  \psfragscanon
  \centering
  \psfrag{a}[c][c][.55]{\hspace{15em} SDoF with fixed $PD$ ($DP$) state~\cite{PRS-15}}
  \psfrag{b}[c][c][.55]{\hspace{18em}\:\:\:SDoF with $DD$ state~(\cite[Theorem 3]{YKPS11}, \eqref{theorem-sdof-miso-bc-achievability-eqautions})}
  \psfrag{c}[c][c][.55]{\hspace{27em}\:\:Achievable SDoF with alternation between $PD$ and $DP$ states~\eqref{theorem-sdof-miso-bc-achievability-eqautions}}
  \psfrag{e}[c][c][.55]{\hspace{12em}\:SDoF/DoF with $PP$ state~\eqref{theorem-sdof-miso-bc-achievability-eqautions}}
  \psfrag{j}[c][c][.8]{\hspace{-1.5em}$(\frac{1}{2},\frac{1}{2})$}
  \psfrag{k}[c][c][.8]{\hspace{1.5em}$(\frac{2}{3},\frac{2}{3})$}
  \psfrag{l}[c][c][.8]{\hspace{-2em}$(1,1)$}
  \psfrag{m}[c][c]{\vspace{1em}$d_1$}
  \psfrag{n}[c][c]{\vspace{1em}$d_2$}
  \includegraphics[width=1\linewidth]{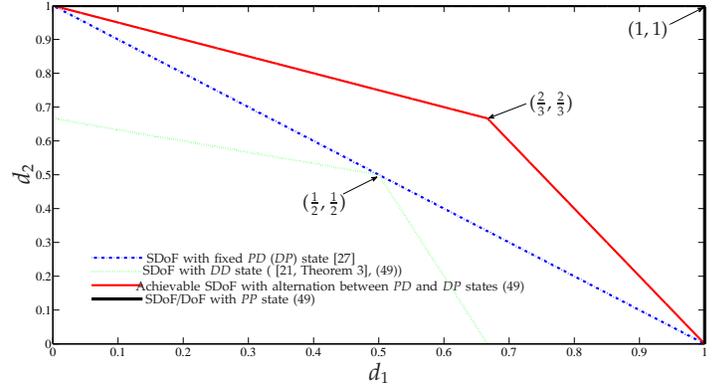}
  \caption{SDoF region of $(2,1,1)$--MISO broadcast channel with alternating CSIT.}
  \psfragscanoff
  \label{MISO-BC-PD-DD-DP-comparison}
  \end{figure}
 
 \vspace{.5em}
 Figure~\ref{MISO-BC-PD-DD-DP-comparison} sheds light on the benefits of alternation between states and shows the SDoF regions of $DD$, $PD$, $PP$ states and the region obtained by alternation between $PD$ and $DP$ states.  It can be easily seen from Figure~\ref{MISO-BC-PD-DD-DP-comparison} that alternation between $PD$ and $DP$ states enlarges the SDoF region in comparison to only $PD$ state. This gain also illustrates the fact that, by joint encoding across these states higher SDoF is achievable.  
 
 \vspace{.5em}
 \begin{remark}[Synergistic Gains in Asymmetric Configurations]
 In Theorem~\ref{theorem-sdof-miso-bc-achievability}, the inner bound provides synergistic benefits of alternating CSIT under \ti{symmetric} assumption of  $\lambda_{PD}=\lambda_{DP}$. Like the model without security constraints~\cite{TJSP13},
 we note that this gain in terms of SDoF is not restricted to symmetric setting and is also preserved under  asymmetric setting, i.e., $\lambda_{PD}\ne \lambda_{DP}$. We consider a simple example in which states $PD$ and $DP$ occur $\lambda_{PD}=1/6$ and $\lambda_{DP}=5/6$ fractions of time, respectively, such that $\lambda_{PD}\ne \lambda_{DP}$ and $\lambda_{PD}+\lambda_{DP}=1$. From~\cite{PRS-15}, it is easy to note that by coding independently over these states, the optimal SDoF is $1$. However, by synergistically using these states one can still obtain higher SDoF as follows. By synergistically using states $PD$ and $DP$ (that gives a SDoF of $4/3$ as will be specified later --- $S_1^{4/3}$ scheme) $1/3$  fraction of time and using $DP$ state in the remaining fraction of time as a separate state, we get
 \begin{eqnarray}
 \text{SDoF}= \frac{1}{3} \times \underbrace{\bigg(\frac{4}{3}\bigg)}_{\mathclap{S_1^{4/3}}} + \frac{2}{3} \times \underbrace{(1)}_{\mathclap{DP}} = \frac{10}{9} \ge 1
 \end{eqnarray}
 which shows the benefits of alternating CSIT under asymmetric configurations.
 \end{remark}
 
 \section{Coding Scheme}\label{schemes}
 In this section, we construct some elemental encoding schemes that provide the main building blocks to establish the inner bound of Theorem~\ref{theorem-sdof-miso-bc-achievability}. 
 
 \subsection{Coding scheme achieving $2$-SDoF}
 The following scheme achieves $2$-SDoF.
 \begin{itemize}
 \item $S^{2}$ -- using $PP$ state, $(d_1,d_2)= (1,1)$ is achievable.
 \end{itemize}
 Due to the availability of perfect CSI of both receivers, the transmitter can zero-force the information leaked to the unintended receiver. Thus, it can be readily shown that one symbol is securely transmitted to each receiver in a single timeslot, yielding 1-SDoF at each receiver.
 
 \subsection{Coding scheme achieving $1$-SDoF}
 The following scheme achieves $1$-SDoF.
 \begin{itemize}
 \item $S^{1}$ -- using $DD$ state, $(d_1,d_2)= (\frac{1}{2},\frac{1}{2})$ is achievable.
 \end{itemize}
 \iffalse
 \begin{table*}
 \centering
 \renewcommand{\arraystretch}{.75}
 \begin{tabular}{| c | c | c | c | c |}
 \hline
 Timeslot & $1$ & $2$ & $3$ & $4$\\ \hline
  $\dv{x}$ & $\dv{u}:=[u_1\:u_2]$ & $ [v_1\: v_2]^T + L_1 (\dv{u})$ & $[w_1\: w_2]^T + M_1 (\dv{u})$ & $\begin{aligned} L_3( w_1,w_2, M_1 (\dv{u}))\\  +M_2( v_1,v_2, L_1 (\dv{u}))\end{aligned}$\\ \hline
 Rx$_1$ & $y_{1,1}= L_1 (\dv{u})$ & $y_{1,2}= L_2( v_1,v_2, L_1 (\dv{u}))$ & $y_{1,3}= L_3( w_1,w_2, M_1 (\dv{u}))$  & $\begin{aligned}y_{1,4}=L_3( w_1,w_2, M_1 (\dv{u}))\\+M_2( v_1,v_2, L_1 (\dv{u}))\end{aligned}$ \\ \hline
 Rx$_2$ & $z_1=  M_1 (\dv{u})$ & $z_2= M_2( v_1,v_2, L_1 (\dv{u}))$ & $z_3=M_3( w_1,w_2, M_1 (\dv{u}))$ & $\begin{aligned}z_4=L_3( w_1,w_2, M_1 (\dv{u})\\+M_2( v_1,v_2, L_1 (\dv{u}))\end{aligned}$\\ \hline
 \end{tabular}
 \vspace{1em}
 \caption{Yang~\ti{et al.} scheme using $DD$ state~\cite{YKPS11}.}
 \label{table}
 \end{table*}
 
 \fi
 For the case in which delayed CSI of both receivers is conveyed to the transmitter, $(d_1,d_2)= (\frac{1}{2},\frac{1}{2})$ SDoF is achievable. The coding scheme in this case is established in \cite{YKPS11} and  we omit it for brevity.
 \subsection{Coding schemes achieving $4/3$-SDoF}
 The following schemes achieve $4/3$ SDoF.
 \begin{enumerate}
 \item $S_1^{4/3}$ -- using $DP$, $PD$ states for $(\frac{1}{2},\frac{1}{2})$ fractions of time, $(d_1,d_2)= (\frac{2}{3},\frac{2}{3})$ SDoF is achievable.
 \item $S_2^{4/3}$ -- using $DD$, $DP$, $PD$ states for $(\frac{1}{3},\frac{1}{3},\frac{1}{3})$ fractions of time, $(d_1,d_2)= (\frac{2}{3},\frac{2}{3})$ SDoF is achievable.
 \end{enumerate}
 \vspace{.5em}
 
 \subsubsection{S$_1^{4/3}$ --- Coding scheme using $DP$ and $PD$ states}
 In the coding scheme that follows, we highlight the benefits of alternation between the states. We now show that by using $PD$ and $DP$ states, $(d_1,d_2)= (\frac{2}{3},\frac{2}{3})$ SDoF is achievable. In this case, transmitter wants to transmit four symbols $(v_1,v_2,v_3,v_4)$ to receiver 1 and wishes to conceal them from receiver 2; and four symbols ($w_1,w_2,w_3,w_4$) to receiver 2 and wishes to conceal them from receiver 1. The communication takes place in six phases, each comprising of only one time slot. In this scheme, the transmitter alternates between different states and chooses $DP$ state at $t=1,3,5$, and $PD$ state at $t= 2,4,6$. In the first phase the transmitter chooses $DP$ state and injects artificial noise, $\tf{u} = [u_{1},u_{2}]^T$. The channel input-output relationship is given by
 \begin{subequations}
 \begin{align}
 y_{1,1} &= {{\tf{h}}}_1 {\tf{u}}, \\
 z_1 &= {{\bf{g}}}_1 {\tf{u}}.
 \end{align}
 \end{subequations}
 At the end of phase 1, the past CSI of receiver 1 is conveyed to the transmitter. In the second phase, utilizing the leverage provided by alternating CSIT  model, the transmitter switches from $DP$ to $PD$ state and sends $\tilde{\tf{v}}:=[v_1, v_2]^T$   along with a linear combination of channel output $y_{1,1}$ of receiver 1 during the first phase. Due to the availability of past CSI of receiver 1 (${{\tf{h}}}_1$) in phase 1 and since the transmitter already knows $\tf{u}$, it can easily re-construct the channel output $y_{1,1}$. During this phase, the transmitter sends
 \begin{eqnarray}
 {\tf{x}}_2 = \left [ v_{1} \:\:  v_{2}\right ]^\text{T} + \left [ y_{1,1} \:\:  \phi  \right ]^\text{T}.
 \end{eqnarray}
 At the end of phase 2, the channel input-output relationship is given by
 \begin{subequations}
 \begin{align}
 \label{phase2-eq-1}
y_{1,2} &= {{\tf{h}}}_2 \tilde{\tf{v}} + h_{21} y_{1,1} , \\
 \label{phase2-eq-2}
 z_2 &= \underbrace{{{\tf{g}}}_2 \tilde{\tf{v}} + g_{21}y_{1,1}}_{\text{interference}}.
 \end{align}
 \end{subequations}
 At the end of phase 2, receiver 2 feeds back the delayed CSI to the transmitter.
 Since receiver 1 knows the CSI $({{\tf{h}}}_2)$ and also the channel output $y_{1,1}$ from phase 1, it subtracts out the contribution of $y_{1,1}$ from the channel output $y_{1,2}$, to obtain one equation with two unknowns ($\tilde{\tf{v}}:=[v_1,v_2]^T$). Thus, receiver 1 requires one extra equation to successfully decode the intended variables, being available as interference or side information at receiver 2.
 
 In the third phase, the transmitter switches from $PD$ to $DP$ state and sends $\tilde{\tf{w}}:=[w_1, w_2]^T$ and $v_3$ along with a linear combination of channel output $z_1$ of receiver 2 during the first phase. The transmitter can easily re-construct $z_1$, since it already knows the perfect CSI $(\tf{g}_1)$ and $\tf{u}$. In phase 3, perfect CSI of receiver 2 (${\tf{g}}_3$) at the transmitter is utilized in two ways, 1) it zero-forces the interference at receiver 2 being caused by symbol $v_3$, and in doing so 2) it also secures symbol $v_3$ which is intended to receiver 1, being eavesdropped by receiver 2. During this phase, the transmitter sends
 \begin{eqnarray}
 {\tf{x}}_3 =   \left [ w_{1} \:\:  w_{2}\right ]^\text{T}  +  \left [ z_{1} \:\:  \phi\right ]^\text{T}  + {\tf{b}_1} v_3, \:\:\:\:\:\:
 \end{eqnarray}
 where $\tf{b}_1 \in \mb{C}^{2\times 1 }$ is the precoding vector chosen such that ${\tf{g}}_3 {\tf{b}_1}$ = 0. At the end of phase 3, the channel input-output relationship is given by
 \begin{subequations}
 \begin{align}
 y_{1,3} &= \underbrace{{{\tf{h}}}_3 \tilde{\tf{w}} + h_{31} z_1}_{\text{interference}} + {{\tf{h}}}_3 {\tf{b}_1} v_3, \\
 z_3 &= {\tf{g}}_3 \tilde{\tf{w}} + g_{31} z_1.
 \end{align}
 \end{subequations}
 At the end of phase 3, receiver 1 feeds back the delayed CSI to the transmitter.
 Receiver 2 can readily subtracts out the contribution of $z_1$ from the channel output $z_3$, to obtain one equation with two unknowns ($\tilde{\tf{w}}:=[w_1,w_2]^T$). Thus, it requires one extra equation to successfully decode the intended variables being available as interference or side information at receiver 1. Receiver 1 gets the intended symbol $v_3$ embedded in with some interference $({{\tf{h}}}_3 \tilde{\tf{w}} + h_{31} z_1)$ from the transmitter. If this interference can be conveyed to the receiver 1, it can then subtracts out the interference's contribution from $y_{1,3}$ and decodes $v_3$ through channel inversion.
 
 At the end of phase 3, due to availability of delayed CSI $(\tf{g}_2, \tf{h}_3)$, the transmitter can learn the interference at receiver 2 in phase 2 and at receiver 1 in phase 3, respectively. In the fourth phase, the transmitter switches from $DP$ to $PD$ state and sends the interference (${\tf{g}}_2 \tilde{\tf{v}} + g_{21} y_{1,1}$) at receiver 2 during the second phase and fresh information $w_3$, where perfect CSI of receiver 1 (${\tf{h}}_4$) is utilized to zero-force the interference being caused by symbol $w_3$ at receiver 1. During this phase, the transmitter sends
 \begin{eqnarray}
 {\tf{x}}_4 =  \left [ {{\tf{g}}}_2 \tilde{\tf{v}} + g_{21} y_{1,1} \:\:  \phi\right ]^\text{T}   + {\tf{b}_2} w_3, \:\:\:\:\:\:
 \end{eqnarray}
 where $\tf{b}_2 \in \mb{C}^{2\times 1 }$ is the precoding vector chosen such that ${\tf{h}}_4 {\tf{b}_2}$ = 0. At the end of phase 4, the channel input-output relationship is given by
 \begin{subequations}
 \begin{align}
y_{1,4} &= h_{41}({\tf{g}}_2 \tilde{\tf{v}} + g_{21} y_{1,1}), \\
 z_4 &= g_{41}({\tf{g}}_2 \tilde{\tf{v}} + g_{21}y_{1,1})+ {{\tf{g}}}_4 {\tf{b}_2} w_3.
 \end{align}
 \end{subequations}
 At the end of phase 4, since receiver 1 knows the CSI and also the channel output $y_{1,1}$ from phase 1, it subtracts out the contribution of $y_{1,1}$ from the channel outputs $(y_{1,2},y_{1,4})$ and decodes ($v_1,v_2$) through channel inversion. Similarly, since receiver 2 knows the CSI and $z_2$ from phase 2, it first subtracts out the contribution of $z_2$ from the channel output $z_4$ and decodes $w_3$ through channel inversion.
 
 In the fifth phase, the transmitter switches from $PD$ to $DP$ state and sends the interference (${{\tf{h}}}_3 \tilde{\tf{w}} + h_{31} z_1$) at receiver 1 during phase 3 and fresh information $v_4$ to receiver 1, where perfect CSI of receiver 2 (${\tf{g}}_5$) is utilized to zero-force the interference being caused by symbol $v_4$ at receiver 2. During this phase, the transmitter sends
 \begin{eqnarray}
 {\tf{x}}_5 = \left [  {{\tf{h}}}_3 \tilde{\tf{w}} + h_{31} z_1 \:\:  \phi\right ]^\text{T}   +  {\tf{b}_3} v_4, \:\:\:\:\:\:
 \end{eqnarray}
 where $\tf{b}_3 \in \mb{C}^{2\times 1 }$ is the precoding vector chosen such that ${\tf{g}}_5 {\tf{b}_3}$ = 0. At the end of phase 5, the channel input-output relationship is given by
 \begin{subequations}
 \begin{align}
 y_{1,5} &= h_{51}( {{\tf{h}}}_3 \tilde{\tf{w}} + h_{31} z_1)+\tf{h}_5 {\tf{b}_3} v_4,\\
 z_5 &= g_{51}( {{\tf{h}}}_3 \tilde{\tf{w}} + h_{31} z_1).
 \end{align}
 \end{subequations}
 At the end of phase 5, since receiver 2 subtracts out the contribution of $z_1$ from the channel outputs $(z_3,z_5)$ and decodes ($w_1,w_2$) through channel inversion.
 Receiver 1 gets the intended symbol $v_4$ embedded within the same interference as in phase 3. If this interference can be conveyed to the receiver 1, it can then subtracts out the interference's contribution from $y_{1,5}$ and decodes $v_4$.
 
 In the sixth phase, the transmitter switches from $DP$ to $PD$ state and sends interference (${{\tf{h}}}_3 \tilde{\tf{w}} + h_{31} z_1$) at receiver 1 during phase 3 with fresh information $w_4$ for receiver 2, where perfect CSI of receiver 1 (${\tf{h}}_6$) is utilized to zero-force the interference being caused by symbol $w_4$ at receiver 1. During this phase the transmitter sends
 \begin{eqnarray}
 {\tf{x}}_6 = \left [{{\tf{h}}}_3 \tilde{\tf{w}} + h_{31} z_1 \:\:  \phi\right ]^\text{T} + {\tf{b}_4} w_4,
 \end{eqnarray}
 where $\tf{b}_4 \in \mb{C}^{2\times 1 }$ is the precoding vector chosen such that ${\tf{h}}_6 {\tf{b}_4}$ = 0. At the end of phase 6, the channel input-output relationship is given by
 \begin{subequations}
 \begin{align}
 y_{1,6} &= h_{61}({{\tf{h}}}_3 \tilde{\tf{w}} + h_{31} z_1),\\
 z_6 &= g_{61}({{\tf{h}}}_3 \tilde{\tf{w}} + h_{31} z_1)+{{\tf{g}}}_6{{\tf{b}}}_4 w_4.
 \end{align}
 \end{subequations}
 At the end of phase 6,  by using $y_{1,6}$ --- receiver 1  subtracts out the contribution of $({{\tf{h}}}_3 \tilde{\tf{w}} + h_{31} z_1)$ from the channel outputs $(y_{1,3},y_{1,5})$ and decodes $v_3$ and $v_4$. Similarly, by using $z_5$ ---  receiver 2   subtracts out the contribution of $({{\tf{h}}}_3 \tilde{\tf{w}} + h_{31} z_1)$ from channel output $z_6$ and decodes $w_4$.
 
 \vspace{.5em}
 \textit{\textbf{Security Analysis.}}
  At the end of phase 6, the channel input-output relationship is given by
 \begin{align}
 {\tf{{y}}} &= \underbrace{ \left [
 \begin{matrix}
 {\tf{h}}_{2} & 0& 0& h_{21} & 0\\
 h_{41}{\tf{g}}_{2} & 0& 0& h_{41}g_{21} & 0\\
 {\tf{0}} & {\tf{h}}_{3} {\tf{b}}_{1} & 0& 0 & 1\\
 {\tf{0}} & 0 & {\tf{h}}_{5} {\tf{b}}_{3} & 0& h_{51}\\
 {\tf{0}} & 0 & 0& 1 & 0\\
 {\tf{0}} & 0 & 0& 0 & h_{61}\\
 \end{matrix} \right ]}_{\tf{H}\in \mb{C}^{6\times 6}} \underbrace{\left [
 \begin{matrix}
 {\tilde{\tf{v}}}\\
 v_3 \\
 v_4\\
 {\tf{h}}_{1} {{\tf{u}}}\\
 {{\tf{h}}}_3 \tilde{\tf{w}} + h_{31} {{\tf{g}}}_1 {\tf{u}}
 \end{matrix} \right ]}_{\mb{C}^{6\times 1}},\\
 \label{main_equation2}
 {\tf{{z}}} &=
 \underbrace{\left[
 \begin{matrix}
 {\tf{g}}_{3} & 0& 0& g_{31} & 0\\
 g_{51}{\tf{h}}_{3} & 0& 0& g_{51}h_{31} & 0\\
 {\tf{0}} & {\tf{g}}_{4} {\tf{b}}_{2} & 0& 0 & g_{41}\\
 g_{61}{\tf{h}}_{3} & 0 & {\tf{g}}_{6} {\tf{b}}_{4} & g_{61}h_{31}& 0\\
 {\tf{0}} & 0 & 0& 1 & 0\\
 {\tf{0}} & 0 & 0& 0 & 1\\
 \end{matrix} \right ]}_{\tf{G}\in \mb{C}^{6\times 6}} \underbrace{\left [
 \begin{matrix}
 {\tilde{\tf{w}}}\\
 w_3 \\
 w_4\\
 {\tf{g}}_{1} {{\tf{u}}}\\
 {{\tf{g}}}_2 \tilde{\tf{v}} + g_{21} {{\tf{h}}}_1 {\tf{u}}
 \end{matrix} \right ]}_{\mb{C}^{6\times 1}}.
 \end{align}
 The information rate to receiver 1 is given by $ I(v_1,v_2,v_3,v_4;\tf{y}|\tf{S}^n)$ and is evaluated as
  \begin{eqnarray}
  &&\hspace{-8em} I(v_1,v_2,v_3,v_4;\tf{y}|\tf{S}^n) \notag\\ &&\hspace{-8em}= I({\tilde{\tf{v}}},v_3,v_4;\tf{{y}}|\tf{S}^n) \notag\\
  && \hspace{-8em} = I({\tilde{\tf{v}}},v_3,v_4,\tf{h}_1 \tf{u}, {{\tf{h}}}_3 \tilde{\tf{w}} + h_{31} z_1;\tf{{y}}|\tf{S}^n) \notag\\
  &&\hspace{-4em} - I(\tf{h}_1 \tf{u}, {{\tf{h}}}_3 \tilde{\tf{w}} + h_{31} z_1;\tf{{y}}|{\tilde{\tf{v}}},v_3,v_4,\tf{S}^n ) \notag\\
  &&\hspace{-8em}   \overset{(a)}{=} \text{rank}(\tf{H})\log (P) - 2\log(P)\notag\\
  && \hspace{-8em}  = 6\log (P) - 2\log(P)\notag\\
  &&\hspace{-8em}   = 4\log(P)
  \end{eqnarray}
 where $(a)$ follows from \cite[Lemma 2]{YKPS11}.
 
 \noindent Similarly, the information leaked to receiver 2 is given by $ I(v_1,v_2,v_3,v_4;\tf{z}|\tf{S}^n)$ and can be bounded as
 \begin{eqnarray}
  && \hspace{-8em} I(v_1,v_2,v_3,v_4;\tf{z}|\tf{S}^n) \notag\\&&\hspace{-8em}  = I({\tilde{\tf{v}}},v_3,v_4;\tf{z}|{\tilde{\tf{w}}},w_3,w_4,\tf{S}^n) \notag\\
  && \hspace{-8em} \le I(\tf{g}_2 \tilde{\tf{v}},v_3,v_4;\tf{z}|{\tilde{\tf{w}}},w_3,w_4,\tf{S}^n)\notag\\
  &&\hspace{-8em} = I(\tf{g}_2 \tilde{\tf{v}},v_3,v_4,\tf{u};\tf{z}|{\tilde{\tf{w}}},w_3,w_4,\tf{S}^n)\notag\\ && - I(\tf{u};\tf{{z}}|\tf{g}_2 \tilde{\tf{v}},{\tilde{\tf{w}}},w_3,w_4,\tf{S}^n)\notag\\
  &&\hspace{-8em} \le I(\tf{g}_2 \tilde{\tf{v}}+g_{21}{{\tf{h}}}_1 {\tf{u}},v_3,v_4,\tf{g}_1\tf{u};\tf{z}|{\tilde{\tf{w}}},w_3,w_4,\tf{S}^n)\notag\\ && - I(\tf{u};\tf{{z}}|\tf{g}_2 \tilde{\tf{v}},{\tilde{\tf{w}}},w_3,w_4,\tf{S}^n)\notag\\
  &&\hspace{-8em} \overset{(a)}{=} 2\log (P) - 2\log(P)\notag\\
 &&\hspace{-8em} = o(\log(P))
\end{eqnarray}
 where $(a)$ follows from~\cite[Lemma 2]{YKPS11}.
 
 \noindent Thus, $4$ symbols are securely transmitted to receiver 1 over a total of $6$ time slots, yielding $d_1= 2/3$ SDoF at receiver 1. Using similar reasoning, it can be readily shown that $4$ symbols are transmitted securely to receiver 2 over $6$ time slots, which yields $d_{2}=2/3$ SDoF at receiver 2.
 \vspace{.5em}
  \begin{remark}
 Notice that, at the end of 6 time slots, $o(\log(P))$ is leaked to the unintended receiver. By combining the above scheme with Wyner's wiretap coding as reasoned in~\cite{YKPS11}, yields the desired SDoF with perfect secrecy.
 \end{remark}
 
 \vspace{.5em}
 \subsubsection{S$_2^{4/3}$ --- Coding scheme using $DD$, $DP$ and $PD$ states}
 In the previous coding scheme S$_1^{4/3}$, availability of \textit{only} delayed CSI of both receivers in the first two phases suffices to achieve $4/3$ SDoF. Thus, by choosing $DD$ state at $t=1,2$, $DP$ state at $t=3,5$, and $PD$ state at $t= 4,6$; $(d_1,d_2)= (\frac{2}{3},\frac{2}{3})$ SDoF pair is achievable.

 \section{Conclusion}\label{conclusion}
In this paper, we studied the SDoF region of a two-receiver channel with an external eavesdropper. We assume that each receiver knows its own CSI and also the past CSI of the other receiver; and, each receiver is allowed to convey either the instantaneous or delayed CSIT. Thus, the overall CSIT vector obtained at the transmitter can alternate between eight possible states. Under these assumptions, we first consider the Gaussian MISO wiretap channel and characterize the full SDoF. 
Next, we consider the general multi-receiver setup and characterize the SDoF region of fixed hybrid states $PPD$, $PDP$, and $DDP$. We then focus our attention on the
symmetric case in which the transmitter is allowed to alternate between $PDD$ and $DPD$ states equal fractions of time and establish bounds on SDoF region. The results established in this work explored the synergistic benefits of alternating CSIT in terms of SDoF; and shows that in comparison to encoding separately over different states, joint encoding across the states provides strictly better secure rates. We also specialized our results to the two-user MISO broadcast  channel. We show that synergistic benefits obtained from alternating CSIT for the multi-receiver channel, can also be harnessed under broadcast setting.

 \appendices
 \section{Proof of Theorem~\ref{theorem-sdof-wt}}
 \label{proof-sdof-wt}

For convenience, we denote the channel output at each receiver as
\begin{align}
& y_1^n := (y^n_{1PP}, y^n_{1PD}, y^n_{1DP}, y^n_{1DD}), \notag\\
& z^n := (z^n_{PP}, z^n_{PD}, z^n_{DP}, z^n_{DD}),\notag
\end{align}
where $y^n_{1S_1S_3} (z^n_{S_1S_3})$ denotes the part of channel output at receiver 1 (eavesdropper), when $(S_1,S_3) \in \{P,D\}^2$ channel state occurs. 

We first introduce a property which will be useful to establish the results in this work. We refer to this as property of \textit{channel output symmetry}~\cite[Lemma 4]{vaze_int}, \cite{MTU14,YKPS11}. We focus our attention to the states $PD$ and $DD$ at the eavesdropper. Recall that for states $PD$ and $DD$ the channel input-output relationship  at the eavesdropper is given by
\begin{eqnarray}
\label{g-chan-2}
{z}_{I,t} = {\tf{g}}_{I,t} \tf{x}_{I,t}+{n}_{I,3t}, \quad\quad\quad  \forall\:\: I \in \{PD,DD\}.
\end{eqnarray}

Next, we consider a statistically indistinguishable receiver which has access to states $PD$ and $DD$, where the channel outputs to this receiver  are
\begin{enumerate}
\item independent from channel outputs at the eavesdropper and
\item identically distributed as the channel outputs at the eavesdropper.
\end{enumerate}
The channel input-output relationship at this statistically indistinguishable receiver at $t$-th time instant is given by
\begin{eqnarray}
\label{g-chan-art}
\tilde{z}_{I,t} = \tilde {\tf{g}}_{I,t} \tf{x}_{I,t}+\tilde{n}_{I,3t}, \quad\quad\quad  \forall\:\: I \in \{PD,DD\}
\end{eqnarray}
where  $\tilde{\tf{g}}_{I,t}$ and ${\tf{g}}_{I,t}$ are identically distributed, and independent of each other --- and all other random variables --- for $I \in \{PD,DD\}$. The additive white Gaussian noise $\tilde{n}_{I,3}$ is assumed to be i.i.d., with  $\tilde{n}_{I,3} \sim \mc{CN}(0,1)$ for $I \in \{PD,DD\}$ and is independent from all random variables. Let $\lambda_{\dv{g}_{I,t}}$ denotes the probability distribution from which, $\tf{g}_{I,t}$ and $\tilde {\tf{g}}_{I,t}$  are independent and identically drawn, for $I \in \{PD,DD\}$.  Let $\dv{S}^n:=\{{\tf{g}}_{I,t} ,\tilde{\tf{g}}_{I,t}\}_{t=1}^n$, for $I \in \{PD,DD\}$.
\vspace{.5em}
\begin{property}
The channel output symmetry states that
\setlength{\arraycolsep}{0.1em}
\begin{align}
\label{local-symmtery}
& h(z_{PD,t},z_{DD,t}|z_{PD}^{t-1},z_{DD}^{t-1},\dv{S}^n)\notag\\& =h(\tilde {z}_{PD,t},\tilde{z}_{DD,t}|z_{PD}^{t-1},z_{DD}^{t-1},\dv{S}^n).
\end{align}
\end{property}
\vspace{.5em}
 
\begin{proof}
We begin the proof as follows.
\setlength{\arraycolsep}{0.001em}
\begin{eqnarray}
&&\hspace{-1.5em} h(z_{PD,t},z_{DD,t}|z_{PD}^{t-1},z_{DD}^{t-1},\dv{S}^n)\notag\\&\overset{(a)}{=}&   \mb {E}_{\lambda_{\dv{g}_{PD,t}},\lambda_{\dv{g}_{DD,t}}}[h(z_{PD,t},z_{DD,t}|z_{PD}^{t-1},z_{DD}^{t-1},\notag\\&&  \dv{g}_{PD,t}=\tf{g}_{PD},\dv{g}_{DD,t}=\tf{g}_{DD},\tilde {\dv{g}}_{PD,t},\tilde {\dv{g}}_{DD,t},\dv{S}^n\setminus\dv{S}_t)] \notag\\
\label{tdenttty2}
&\overset{(b)}{=}&  \mb {E}_{\lambda_{\dv{g}_{PD,t}},\lambda_{\dv{g}_{DD,t}}}[ h({\tf{g}}_{PD} \tf{x}_{PD,t}+{n}_{PD,3t},{\tf{g}}_{DD} \tf{x}_{DD,t}+{n}_{DD,3t}|\notag\\&& z_{PD}^{t-1},z_{DD}^{t-1},  \dv{S}^n\setminus\dv{S}_t)] \notag\\
&\overset{(c)}{=}&   \mb {E}_{\lambda_{\dv{g}_{PD,t}},\lambda_{\dv{g}_{DD,t}}}[h({\tf{g}}_{PD} \tf{x}_{PD,t}+\tilde{n}_{PD,3t},{\tf{g}}_{DD} \tf{x}_{DD,t}+\tilde{n}_{DD,3t}|\notag\\&&z_{PD}^{t-1},z_{DD}^{t-1},  \dv{S}^n\setminus\dv{S}_t)] \notag\\
&\overset{(d)}{=}&   \mb {E}_{\lambda_{\dv{g}_{PD,t}},\lambda_{\dv{g}_{DD,t}}}[ h({\tf{g}}_{PD} \tf{x}_{PD,t}+\tilde{n}_{PD,3t},{\tf{g}}_{DD} \tf{x}_{DD,t}+\tilde{n}_{DD,3t}|\notag\\&& z_{PD}^{t-1},z_{DD}^{t-1}, \tilde {\dv{g}}_{PD,t}=\tf{g}_{PD},\tilde {\dv{g}}_{DD,t}=\tf{g}_{DD}, \dv{S}^n\setminus\dv{S}_t)] \notag\\  
&\overset{(e)}{=}&  \mb {E}_{\lambda_{\dv{g}_{PD,t}},\lambda_{\dv{g}_{DD,t}}}[ h(\tilde{z}_{PD,t},\tilde{z}_{DD,t}|z_{PD}^{t-1},z_{DD}^{t-1},\notag\\&& \tilde {\dv{g}}_{PD,t}=\tf{g}_{PD},\tilde {\dv{g}}_{DD,t}=\tf{g}_{DD},  {\dv{g}}_{PD,t}, {\dv{g}}_{DD,t},\dv{S}^n\setminus\dv{S}_t)] \notag\\
&=&    h(\tilde{z}_{PD,t},\tilde{z}_{DD,t}|z_{PD}^{t-1},z_{DD}^{t-1},\dv{S}^n)
\end{eqnarray}
\noindent where $(a)$ follows due to the definition of differential entropy, $(b)$ follows because $\tf{x}_{I,t}$ is independent from $( {\tf{g}}_{I,t}, \tilde {\tf{g}}_{I,t})$, $(c)$ follows because ${n}_{I,3t}$ and $\tilde{n}_{I,3t}$ are independent from all other random variables and have same statistics, $(d)$ and $(e)$ follow because since $({\tf{g}}_{I,t},\tilde {\tf{g}}_{I,t})$  belong to same distribution  $\lambda_{\tf{g}_{I,t}}$ and due to the independence of $\tf{x}_{I,t}$ and $({\tf{g}}_{I,t},\tilde {\tf{g}}_{I,t})$; for $I \in \{PD,DD\}$.
\end{proof}
\vspace{.5em}

Before proceeding to state the proof of Theorem~\ref{theorem-sdof-wt}, we first digress to provide a useful lemma which we will repetitively use in this work.
  \vspace{.5em}
 \begin{lemma}
 \label{lemma}
 For the Gaussian MISO channel in~\eqref{g-chan-mu} and~\eqref{g-chan}, following inequalities hold
 \begin{subequations}
 \begin{align}
 \label{l1}
 2 h(z^n_{PD}, z^n_{DP}, z^n_{DD}|\dv{S}^n) &\dot{\geq}  h(y^n_{1PD}, y^n_{1DD}| \dv{S}^n) ,\\
 \label{l2}
 2 h(y^n_{1PD}, y^n_{1DP}, y^n_{1DD}|\dv{S}^n) &\dot{\geq}  h(z^n_{DP}, z^n_{DD}| \dv{S}^n) ,\\
 \label{l3}
h(z^n_{PD}, z^n_{DP}, z^n_{DD}|\dv{S}^n) &\dot{\geq}  h(y^n_{1PD}, y^n_{1DD}|z^n_{PD}, z^n_{DP}, z^n_{DD}, \dv{S}^n) ,\\
 \label{l4}
  h(y^n_{1PD}, y^n_{1DP}, y^n_{1DD}|\dv{S}^n) &\dot{\geq}  h( z^n_{DP}, z^n_{DD}|y^n_{1PD} , y^n_{1DP}, y^n_{1DD}, \dv{S}^n).
 \end{align}
 \end{subequations}
 \end{lemma}
   
  \begin{IEEEproof}
 \noindent We now provide the proof of~\eqref{l1} and~\eqref{l3}; due to the symmetry the rest of the inequalities follow straightforwardly.
 
 We begin the proof as follows.
  \setlength{\arraycolsep}{0.1em}
\begin{align}
  & 2 h(z^n_{PD}, z^n_{DP}, z^n_{DD}|\dv{S}^n)\notag\\
&=   2 h(z^n_{PD},  z^n_{DD}|\dv{S}^n)+2 h(z^n_{DP}|z^n_{PD},z^n_{DD},\dv{S}^n)    \notag\\
&\overset{(f)}{\geq}   2 h(z^n_{PD},  z^n_{DD}|\dv{S}^n)+ h(z^n_{DP}|z^n_{PD},z^n_{DD},\dv{S}^n)\notag\\&\:\:\:\:\: + \underbrace{h(z^n_{DP}|z^n_{PD},z^n_{DD},\dv{x}^n,\dv{S}^n)}_{ \le no(\log(P))}    \notag\\
& \overset{(g)}{\geq}  2 \sum_{t=1}^n h(z_{PD,t},  z_{DD,t}|z^{t-1}_{PD},  z^{t-1}_{DD},\dv{S}^n )    +  h(z^n_{DP}|z^n_{PD},z^n_{DD},\dv{S}^n )  \notag\\  
& \overset{(h)}{=} \sum_{t=1}^n h(z_{PD,t}, z_{DD,t}|z^{t-1}_{PD},  z^{t-1}_{DD},\dv{S}^n)\notag\\&\:\:\:\:\:+h(\tilde{z}_{PD,t}, \tilde{z}_{DD,t}|z^{t-1}_{PD}, z^{t-1}_{DD},\dv{S}^n) +  h(z^n_{DP}|z^n_{PD},z^n_{DD},\dv{S}^n)\notag\\   
&\geq  \sum_{t=1}^n h(z_{PD,t}, z_{DD,t},\tilde{z}_{PD,t}, \tilde{z}_{DD,t}|z^{t-1}_{PD},  z^{t-1}_{DD},\dv{S}^n)\notag\\&\:\:\:\:\: +  h(z^n_{DP}|z^n_{PD},z^n_{DD},\dv{S}^n) \notag\\  
&= \sum_{t=1}^n h(z_{PD,t}, z_{DD,t},\tilde{z}_{PD,t}, \tilde{z}_{DD,t},y_{1PD,t}, y_{1DD,t} |z^{t-1}_{PD}, z^{t-1}_{DD},\dv{S}^n)\notag\\&\:\:\:\:\:-h(y_{1PD,t}, y_{1DD,t} |z^{n}_{PD}, z^{n}_{DD},\tilde{z}_{PD,t}, \tilde{z}_{DD,t},\dv{S}^n)\notag\\&\:\:\:\:\:+  h(z^n_{DP}|z^n_{PD},z^n_{DD},\dv{S}^n)    \notag\\
&= \sum_{t=1}^n h(z_{PD,t}, z_{DD,t}, y_{1PD,t}, y_{1DD,t} |z^{t-1}_{PD}, z^{t-1}_{DD},\dv{S}^n)\notag\\&\:\:\:\:\:+\underbrace{h(\tilde{z}_{PD,t}, \tilde{z}_{DD,t} |z^{n}_{PD}, z^{n}_{DD},y_{1PD,t}, y_{1DD,t},\dv{S}^n)}_{ = no(\log(P))}\notag\\&\:\:\:\:\:-\underbrace{h(y_{1PD,t}, y_{1DD,t} |z^{n}_{PD}, z^{n}_{DD},\tilde{z}_{PD,t}, \tilde{z}_{DD,t},\dv{S}^n)}_{ = no(\log(P))}\notag\\&\:\:\:\:\:+  h(z^n_{DP}|z^n_{PD},z^n_{DD},\dv{S}^n)    \notag\\
&\overset{(i)}{\geq} \sum_{t=1}^n h(z_{PD,t}, z_{DD,t},y_{1PD,t}, y_{1DD,t} |z^{t-1}_{PD}, z^{t-1}_{DD},y^{t-1}_{1PD}, y^{t-1}_{1DD}, \dv{S}^n)\notag\\&\:\:\:+  h(z^n_{DP}|z^n_{PD},z^n_{DD},\dv{S}^n) +no(\log(P))\notag\\
\label{term-int-2}
&= h(z^n_{PD}, z^n_{DD},y^n_{1PD}, y^n_{1DD}| \dv{S}^n)+  h(z^n_{DP}|z^n_{PD},z^n_{DD},\dv{S}^n)\notag\\&\:\:\:+no(\log(P)) \notag\\
&\overset{(j)}{\geq}   h(z^n_{PD}, z^n_{DD},y^n_{1PD}, y^n_{1DD}| \dv{S}^n)\notag\\&\:\:\:\:\:+  h(z^n_{DP}|z^n_{PD},z^n_{DD},y^n_{PD}, y^n_{1DD},\dv{S}^n)+no(\log(P)) \notag\\
&=  h(z^n_{PD}, z^n_{DP}, z^n_{DD},y^n_{1PD}, y^n_{1DD}| \dv{S}^n)+no(\log(P))\\
&\geq  h(y^n_{1PD}, y^n_{1DD}| \dv{S}^n)+no(\log(P))
\end{align}
\noindent where $(f)$ and $(j)$ follow from the fact that conditioning reduces entropy, $(g)$ follows because given $(\tf{x}^n,\tf{S}^n)$, $ {z}^n_{DP}$ can be recovered within bounded noise distortion, $(h)$ follows from the property of channel output symmetry~\eqref{local-symmtery}, $(i)$ follows from the fact that conditioning reduces entropy; and, $(\tilde{z}_{PD,t}, \tilde{z}_{DD,t})$ and $(y_{PD,t}, y_{DD,t})$ can be reconstructed within bounded noise distortion form $(z^{n}_{PD}, z^{n}_{DD},y_{1PD,t}, y_{1DD,t},\tf{S}^n$), and $(z^{n}_{PD}, z^{n}_{DD},\tilde{z}_{PD,t}, \tilde{z}_{DD,t},\tf{S}^n)$, respectively.
 
We can also bound the term in~\eqref{term-int-2} as follows. Continuing from~\eqref{term-int-2}, we get
\begin{align}
\label{lemma-term2}
2 h(z^n_{PD}, z^n_{DP}, z^n_{DD}|\tf{S}^n)& \geq  h(z^n_{PD}, z^n_{DP}, z^n_{DD},y^n_{1PD}, y^n_{1DD}| \tf{S}^n)\notag\\&\:\:\: +no(\log(P)) \notag\\
&=  h(z^n_{PD}, z^n_{DP}, z^n_{DD}| \tf{S}^n)\notag\\&\:\:\:+  h(y^n_{1PD}, y^n_{1DD}|z^n_{PD}, z^n_{DP}, z^n_{DD},\tf{S}^n)\notag\\&\:\:\:+no(\log(P)).
\end{align}
From~\eqref{lemma-term2}, it implies that
\begin{align}
h(z^n_{PD}, z^n_{DP}, z^n_{DD}|\tf{S}^n)& \geq  (y^n_{1PD}, y^n_{1DD}|z^n_{PD}, z^n_{DP}, z^n_{DD},\tf{S}^n)\notag\\&\:\:\:+no(\log(P)).
\end{align}
 This concludes the proof.
 \end{IEEEproof}

 \vspace{.5em}
 \textit{\textbf{\underline{Achievability}.}} We first sketch some elemental coding schemes that are used to establish the achievability in Theorem~\ref{theorem-sdof-wt}.  
 
 \vspace{.5em}
 \textit{$S^{1}$---\textbf{Coding schemes achieving $1$-SDoF:}}
 For $PP$, and $DP$ states $1$-SDoF is achievable. Due to the availability of perfect CSI of the unintended receiver (wire-taper), the transmitter can zero-force the information leaked to it. Thus, it can be readily shown that one symbol is securely transmitted to the legitimate receiver in a single timeslot, yielding 1-SDoF.
 
 For the case in which $PD$ state occurs, the scheme follows straightforwardly from the one described in subsection~\ref{pd1} by removing the receiver 2,  yielding 1-SDoF.
 \vspace{.5em}
 
 \textit{$S^{2/3}$---\textbf{Coding scheme achieving $2/3$-SDoF:}}
 For the case in which $DD$ state occurs, $2/3$ SDoF is achievable. The coding scheme in this case is similar to the one in~\cite[Section IV-B-2]{YKPS11} for the wiretap channel with delayed CSIT from both receivers and is omitted for brevity.

 The achievable SDoF of Theorem~\ref{theorem-sdof-wt} then follows by choosing $PP, PD,DP$ and $DD$ states $\lambda_{PP}$, $\lambda_{PD}$, $\lambda_{DP}$ and $\lambda_{DD}$ fractions of time, respectively, which yields
 \begin{align}
 d_s & = \lambda_{PP} (1)+ \lambda_{PD} (1) + \lambda_{DP} (1) + \lambda_{DD} \big(\frac{2}{3}\big)\notag\\
 &= 1-\frac{ \lambda_{DD}}{3}.
 \end{align}

 \textit{\textbf{\underline{Converse Proof}.}} The details of the converse proof appears in~\cite{AZS-15}.

 \section{Proof of Theorem~\ref{theorem-sdof-mu-ppd}}
 \label{proof-sdof-mu-ppd-converse}
 The proof of~\eqref{eq-ppda} and~\eqref{eq-ppdb} follows along similar lines as in the proof of Theorem~\ref{theorem-sdof-wt}. In what follows, we provide the proof of~\eqref{eq-ppdc}.
 
  We begin the proof as follows.
   \setlength{\arraycolsep}{0.05em}
 \begin{align}
  \label{sum-a}
 & n(R_1+R_2)\notag\\&= {H}(W_1,W_2|z^n,\tf{S}^n) \\
 &= {H}(W_1,W_2|\tf{S}^n)-I(W_1,W_2;z^n|\tf{S}^n)\notag \\
 &= {H}(W_1|\tf{S}^n)+{H}(W_2|W_1,\tf{S}^n)-I(W_1,W_2;z^n|\tf{S}^n)\notag \\
 &= H(W_1|y_1^n,\tf{S}^n)+I(W_1;y_1^n|\tf{S}^n)+  {H}(W_2|W_1,y_{2}^n,\tf{S}^n)\notag \\&\:\:\:+I(W_2;y_2^n|W_1,\tf{S}^n)-I(W_1,W_2;z^n|\tf{S}^n)\notag \\
 &\overset{(a)}{\le} I(W_1;y_1^n|\tf{S}^n)+   I(W_2;y_2^n|W_1,\tf{S}^n)-I(W_1,W_2;z^n|\tf{S}^n)+n\epsilon_{n}  \notag\\
 & \le I(W_1;y_1^n,z^n|\tf{S}^n)+   I(W_2;y_2^n,z^n|W_1,\tf{S}^n)-I(W_1,W_2;z^n|\tf{S}^n)\notag \\&\:\:\:+n\epsilon_{n}  \notag\\
 \label{sum-int}
 &= I(W_1;y_1^n|z^n,\tf{S}^n)+ I(W_2;y_2^n|z^n, W_1,\tf{S}^n)+n\epsilon_{n}  \\
 &\le I(W_1;y_1^n, \tilde{z}^n|z^n,\tf{S}^n)+ I(W_2;y_2^n,\tilde{z}^n|z^n, W_1,\tf{S}^n)+n\epsilon_{n}  \notag\\
   &= h(\tilde{z}^n|z^n,\tf{S}^n)+\underbrace{h(y_1^n |\tilde{z}^n,z^n,\tf{S}^n)}_{ \le no(\log(P))}-h(y_1^n, \tilde{z}^n|z^n,W_1,\tf{S}^n)\notag\\&\:\:\:+h(\tilde{z}^n|z^n, W_1,\tf{S}^n) +\underbrace{h(y_2^n|\tilde{z}^n,z^n, W_1,\tf{S}^n))}_{ \le no(\log(P))}\notag\\&\:\:\:-h(y_2^n,\tilde{z}^n|z^n, W_1,W_2,\tf{S}^n)+n\epsilon_{n}  \notag\\ 
&\overset{(b)}{\le} h(\tilde{z}^n|z^n,\tf{S}^n)+h(\tilde{z}^n|z^n, W_1,\tf{S}^n)+no(\log(P))+n\epsilon_{n}  \notag\\ 
&=\sum_{t=1}^{n}h(\tilde{z}_t|\tilde{z}_{t-1},z^n,\tf{S}^n)+h(\tilde{z}_t|\tilde{z}_{t-1},z^n, W_1,\tf{S}^n)+no(\log(P))\notag\\&\:\:\:+n\epsilon_{n}  \notag\\  &\overset{(c)}{=}\sum_{t=1}^{n}h(z_t|\tilde{z}_{t-1},z^n,\tf{S}^n)+h(z_t|\tilde{z}_{t-1},z^n, W_1,\tf{S}^n)+no(\log(P))\notag\\&\:\:\:+n\epsilon_{n}  \notag\\  
&\le\sum_{t=1}^{n}h(z_t|z^{t-1},\tf{S}^n)+h(z_t|z_{t-1}, W_1,\tf{S}^n)+no(\log(P))+n\epsilon_{n}  \notag\\   
&= h(z^n|\tf{S}^n)+h(z^n| W_1,\tf{S}^n)+no(\log(P))+n\epsilon_{n}  \notag\\   
 \label{term-mu-ppd}  
&\le 2n\log(P)+no(\log(P))+n\epsilon_{n} 
 \end{align}
  \noindent where $\epsilon_n \rightarrow 0$ as $n \rightarrow \infty$; $(a)$ follows from Fano's inequality, $(b)$ follows  because $y_{1}^n$ and $y_{2}^n$  can be obtained within bounded noise distortion form $(z^n, \tilde{z}^n,\tf{S}^n)$, $(c)$ follows due to the property of channel output symmetry~\eqref{local-symmtery}.
  
 Then, dividing both sides of~\eqref{term-mu-ppd} by $n\log(P)$ and taking $\lim {P \rightarrow \infty}$ and $\lim {n \rightarrow \infty} $, we get
 \begin{equation}
 d_1+d_2 \le 2.
 \end{equation}
  This concludes the proof.

 \section{Proof of Theorem~\ref{theorem-sdof-mu-pdp}}
 \label{proof-sdof-mu-pdp-converse}
 The proof of~\eqref{eq-pdpa} follows along similar lines as in the proof of Theorem~\ref{theorem-sdof-wt} and is omitted for brevity.  We now provide the proof of~\eqref{eq-pdpb}.
   \setlength{\arraycolsep}{0.1em}
 \begin{align}
 \label{term-mu-pdp1}
& n(R_1+R_2)\notag\\&= {H}(W_1,W_2|z^n,\tf{S}^n)\notag \\
 &= {H}(W_1,W_2|\tf{S}^n)-I(W_1,W_2;z^n|\tf{S}^n)\notag \\
 &= {H}(W_2|\tf{S}^n)+{H}(W_1|W_2,\tf{S}^n)-I(W_1,W_2;z^n|\tf{S}^n)\notag \\
 &= H(W_2|y_2^n,\tf{S}^n)+I(W_2;y_2^n|\tf{S}^n)  +{H}(W_1|W_2,y_{1}^n,\tf{S}^n)\notag \\&\:\:\:+I(W_1;y_1^n|W_2,\tf{S}^n)-I(W_1,W_2;z^n|\tf{S}^n)\notag \\
 &\overset{(a)}{\le} I(W_2;y_2^n|\tf{S}^n)+   I(W_1;y_1^n|W_2,\tf{S}^n)-I(W_1,W_2;z^n|\tf{S}^n)+n\epsilon_{n}  \notag\\
 & \le I(W_2;y_2^n,z^n|\tf{S}^n)+   I(W_1;y_1^n,z^n|W_2,\tf{S}^n)-I(W_1,W_2;z^n|\tf{S}^n)\notag \\&\:\:\:+n\epsilon_{n}  \notag\\
 &= I(W_2;y_2^n|z^n,\tf{S}^n)+ I(W_1 ;y_1^n|z^n, W_2,\tf{S}^n)+n\epsilon_{n}  \notag\\
  &\le I(W_2;y_2^n|z^n,\tf{S}^n)+ I(W_1 ;y_1^n,y_2^n|z^n, W_2,\tf{S}^n)+n\epsilon_{n}  \notag\\
 &\le h(y_2^n|\tf{S}^n)+h(y_2^n|z^n,W_2,\tf{S}^n)+\underbrace{h(y_1^n|z^n,y_2^n, W_2,\tf{S}^n)}_{ \le no(\log(P))}+n\epsilon_{n}\notag\\
 &\overset{(b)}{\le} h(y_2^n|\tf{S}^n)+h(y_2^n|W_2,\tf{S}^n)+no(\log(P))+n\epsilon_{n} 
 \end{align}
  \noindent where $\epsilon_n \rightarrow 0$ as $n \rightarrow \infty$; $(a)$ follows from Fano's inequality, and $(b)$ follows  because $y_{1}^n$ can be obtained within bounded noise distortion form $(z^n, y_2^n,\tf{S}^n)$.
  
We can also bound $R_2$ as follows.
\begin{eqnarray}
 \label{term-mu-pdp2}
 nR_2&=& H(W_2|\tf{S}^n)\notag\\
 &=& I(W_2;y_2^n|\tf{S}^n)+H(W_2|y_2^n,\tf{S}^n)\notag\\
&\overset{(c)}{\le}& h(y_2^n|\tf{S}^n)-h(y_2^n|W_2,\tf{S}^n)+n\epsilon_n
 \end{eqnarray}
where $(c)$ follows from Fano's inequality. 
Combining~\eqref{term-mu-pdp1} and~\eqref{term-mu-pdp2}, we get 
\begin{eqnarray}
 \label{term-mu-f}
 n(R_1+2R_2)&\le& 2h(y_2^n|\tf{S}^n)+no(\log(P))+n\epsilon_{n} \notag\\
 &\le& 2n\log(P)+no(\log(P))+n\epsilon_{n}.
 \end{eqnarray}

Then, dividing both sides of~\eqref{term-mu-f} by $n\log(P)$ and taking $\lim {P \rightarrow \infty}$ and $\lim {n \rightarrow \infty} $, we get
 \begin{eqnarray}
 d_1+2d_2 &\le& 2.
 \end{eqnarray}
 
 This concludes the proof.

\section{Proof of Theorem~\ref{theorem-sdof-mu-main-outer-bound}}
\label{proof-sdof-mu-outer-bound}
We denote the channel output at each receiver as
\begin{align}
& y_1^n := (y_{1,PDD}^n, y_{1,DPD}^n), \notag\\
& y_2^n := (y_{2,DPD}^n, y_{2,DPD}^n), \notag\\
& z^n := (z^n_{PDD}, z^n_{DPD}),
\end{align}
where $y_{i,S_1S_2S_3}^n, z^n_{S_1S_2S_3}$ denotes the part of channel output at receiver  $i\in\{1, 2\}$ and eavesdropper, when $(S_1,S_2,S_3) \in \{PDD,DPD\}$ channel state occurs. Next, we provide a Lemma which will be used extensively to establish the outer bound.
 \vspace{.5em}
 \begin{lemma}
 \label{lemma-mu}
 For the Gaussian multi-user wiretap channel in~\eqref{g-chan-mu}, following inequalities hold
  \begin{subequations}
 \begin{align}
 \label{l1u}
 2 h(y_1^{n}|\tf{S}^n) &\dot{\geq}  h(y_1^n, y_{2DPD}^{n}| \tf{S}^n),\\
 \label{l1uo}
  2 h(y_2^{n}|\tf{S}^n) &\dot{\geq}  h(y_2^n, y_{1PDD}^{n}| \tf{S}^n),\\
 \label{l2u}
2 h(y_1^{n}|\tf{S}^n) &\dot{\geq}  h(z^n_{DPD}| \tf{S}^n),\\  
 \label{l2uo}
2 h(y_2^{n}|\tf{S}^n) &\dot{\geq}  h(z^n_{PDD}| \tf{S}^n),\\ 
\label{l4u}
2 h(z^{n}|\tf{S}^n) &\dot{\geq}  h(y_1^n,z^n| \tf{S}^n),\\  
 \label{l4uo}
2 h(z^{n}|\tf{S}^n) &\dot{\geq}  h(y_2^n,z^n| \tf{S}^n),\\  
 \label{l3u}
2 h(y_1^n,z^{n}|\tf{S}^n) &\dot{\geq}  h(y_1^n, z^n, y_{2DPD}^{n}| \tf{S}^n) ,\\
 \label{l3uo}
2 h(y_2^n,z^{n}|\tf{S}^n) &\dot{\geq}  h(y_1^n, z^n, y_{2PDD}^{n}| \tf{S}^n).
 \end{align}
 \end{subequations}
 \end{lemma}
\vspace{.5em}
\begin{proof}
The proof of~\eqref{l1u}-\eqref{l4uo} follows along similar lines as in Lemma~\ref{lemma}.  In what follows, we now provide the proof of~\eqref{l3u}. Due to the symmetry, the proof of~\eqref{l3uo} follows along similar lines and is omitted.\\ We begin the proof as follows.
 \setlength{\arraycolsep}{0.2em}
 \begin{align}
&  2 h(y_1^n, z^n|\tf{S}^n)\notag\\
& =   2 h(y_{1DPD}^n, z^n|\tf{S}^n)+2 h(y_{1PDD}^n|y_{1DPD}^n, z^n,\tf{S}^n)    \notag\\
&\geq 2 h(y_{1DPD}^n, z^n|\tf{S}^n)+ h(y_{1PDD}^n|y_{1DPD}^n, z^n,\tf{S}^n)\notag\\&+ \underbrace{h(y_{1PDD}^n|y_{1DPD}^n, \tf{x}^n, z^n,\tf{S}^n)}_{ \le no(\log(P))}    \notag\\
& \overset{(a)}{\geq}  \sum_{t=1}^n h(y_{1DPD,t}, z_t|y_{1DPD}^{t-1}, z^{t-1},\tf{S}^n)\notag\\&+ h(y_{1DPD,t}, z_t|y_{1DPD}^{t-1}, z^{t-1},\tf{S}^n)+h(y_{1PDD}^n|y_{1DPD}^n, z^n,\tf{S}^n)  \notag\\  
& \overset{(b)}{=}    \sum_{t=1}^n h(y_{1DPD,t}, z_t|y_{1DPD}^{t-1}, z^{t-1},\tf{S}^n)\notag\\&+ h(\tilde{y}_{1DPD,t}, \tilde{z}_t|y_{1DPD}^{t-1}, z^{t-1},\tf{S}^n)+h(y_{1PDD}^n|y_{1DPD}^n, z^n,\tf{S}^n)  \notag\\ 
&{\geq}  \sum_{t=1}^n h(y_{1DPD,t}, \tilde{y}_{1DPD,t}, z_t, \tilde{z}_t|y_{1DPD}^{t-1}, z^{t-1},\tf{S}^n)\notag\\&+h(y_{1PDD}^n|y_{1DPD}^n, z^n,\tf{S}^n)  \notag\\ 
& {=}  \sum_{t=1}^n h(y_{1DPD,t}, \tilde{y}_{1DPD,t}, y_{2DPD,t}, z_t, \tilde{z}_t|y_{1DPD}^{t-1}, z^{t-1},\tf{S}^n)\notag\\&-h(y_{2DPD,t}|y_{1DPD}^n, \tilde{y}_{1DPD,t}, z^n, \tilde{z}_t,\tf{S}^n ) + h(y_{1PDD}^n|y_{1DPD}^n, z^n,\tf{S}^n)  \notag\\ 
& {\geq}  \sum_{t=1}^n h(y_{1DPD,t},  y_{2DPD,t}, z_t|y_{1DPD}^{t-1}, z^{t-1},\tf{S}^n)\notag\\&+\underbrace{h(\tilde{y}_{1DPD,t}| y_{1DPD}^n, y_{2DPD,t}, z^{n},\tf{S}^n)}_{ = no(\log(P))}\notag\\&-\underbrace{h(y_{2DPD,t}|y_{1DPD}^n, \tilde{y}_{1DPD,t}, z^n, \tilde{z}_t,\tf{S}^n )}_{ = no(\log(P))}+ h(y_{1PDD}^n|y_{1DPD}^n, z^n,\tf{S}^n)  \notag\\
&  \overset{(c)}{\geq} \sum_{t=1}^n h(y_{1DPD,t}, y_{2DPD,t}, z_t, |y_{1DPD}^{t-1},y_{2DPD}^{t-1}, z^{t-1},\tf{S}^n)\notag\\&+ h(y_{1PDD}^n|y_{1DPD}^n, z^n,\tf{S}^n)+no(\log(P))\notag\\   
&=  h(y_{1DPD}^n, y_{2DPD}^n, z^n|\tf{S}^n)+ h(y_{1PDD}^n|y_{1DPD}^n, z^n,\tf{S}^n)\notag\\&+no(\log(P))\notag\\ 
&\geq h(y_{1DPD}^n, y_{2DPD}^n, z^n|\tf{S}^n)+ h(y_{1PDD}^n|y_{1DPD}^n, y_{2DPD}^n, z^n,\tf{S}^n)\notag\\&+no(\log(P))\notag\\ 
&=  h(y_{1DPD}^n, y_{1PDD}^n, y_{2DPD}^n, z^n|\tf{S}^n) +no(\log(P))\notag\\
&=  h(y_1^n, z^n,y_{2DPD}^n| \tf{S}^n)+no(\log(P))
\end{align}
\noindent where $(a)$ follows because given $(\tf{x}^n,\tf{S}^n)$, $ y_{1PDD}^n$ can be recovered within bounded noise distortion, $(b)$ follows by invoking the property of channel output symmetry~\eqref{local-symmtery} to the multi-user wiretap channel, $(c)$ follows from the fact that conditioning reduces entropy; and, $\tilde{y}_{1DPD,t}$ and $ y_{2DPD,t}$ can be reconstructed within bounded noise distortion form $(y_{1DPD}^n, y_{2DPD,t}, z^{n},\tf{S}^n)$, and $(y_{1DPD}^n, \tilde{y}_{1DPD,t}, z^n, \tilde{z}_t,\tf{S}^n)$, respectively.
\end{proof}
\vspace{1em}

We now provide the proof of~\eqref{miso-mu-outer-1}. We begin the proof as follows.
\setlength{\arraycolsep}{0.2em}
  %======== SINGLE COLUMN ===============
\begin{align}
\label{term-1-mu}
& nR_{1}\notag\\&= {H}(W_1 | z^n )\notag \\
&\overset{(a)}{\le} I(W_1; y_1^n | \tf{S}^n)- I(W_1; z^n| \tf{S}^n)+n\epsilon_{n} \notag\\
&= h( y_1^n| \tf{S}^n)-h( y_1^n| W_1, \tf{S}^n)- h(z^n| \tf{S}^n)+ h(z^n |W_1, \tf{S}^n)+n\epsilon_{n}\notag \\
&\overset{(b)}{\le} h( y_1^n| \tf{S}^n)-\frac{1}{2}h( z^n_{DPD} | W_1, \tf{S}^n)- h(z^n_{PDD}, z^n_{DPD}| \tf{S}^n)\notag\\&+ h(z^n_{PDD}, z^n_{DPD}|W_1, \tf{S}^n)+n\epsilon_{n}\notag \\
&= h( y_1^n| \tf{S}^n)-\frac{1}{2}h( z^n_{DPD}| W_1, \tf{S}^n)- h( z^n_{DPD}| \tf{S}^n)\notag\\&-h(z^n_{PDD}| z^n_{DPD}, \tf{S}^n)+h( z^n_{DPD}|W_1, \tf{S}^n)\notag\\&+h(z^n_{PDD}|z^n_{DPD},W_1, \tf{S}^n)+n\epsilon_{n}\notag \\
&= h( y_1^n| \tf{S}^n)+\frac{1}{2}h( z^n_{DPD}| W_1, \tf{S}^n)- h( z^n_{DPD}| \tf{S}^n)\notag\\&-h(z^n_{PDD}| z^n_{DPD}, \tf{S}^n)+h(z^n_{PDD}|z^n_{DPD},W_1, \tf{S}^n)+n\epsilon_{n}\notag \\
&\overset{(c)}{\le} h( y_1^n| \tf{S}^n)+\frac{1}{2}h( z^n_{DPD}|  \tf{S}^n)- h( z^n_{DPD}| \tf{S}^n)-h(z^n_{PDD}| z^n_{DPD}, \tf{S}^n)\notag\\&+h(z^n_{PDD}|z^n_{DPD}, \tf{S}^n)+n\epsilon_{n}\notag \\
&= h( y_1^n| \tf{S}^n)-\frac{1}{2}h( z^n_{DPD}|  \tf{S}^n)+n\epsilon_{n}
 \end{align}
\noindent where $(a)$ follows from Fano's inequality, $(b)$ follows by applying~\eqref{l2u} with conditioning over $W_1$, and $(c)$ follows due to the fact that conditioning reduces entropy.
  
 \noindent We can also bound $R_1$ as follows.
 \begin{eqnarray}
 \label{term-2-mu}
 nR_{1}&=& {H}(W_1|\tf{S}^n)\notag \\
 &=& I(W_1;y_1^n|\tf{S}^n)+{H}(W_1|y_1^n, \tf{S}^n)\notag \\
 &\overset{(b)}{\le}& h(y_1^n | \tf{S}^n)- \frac{1}{2}h(y^n_1,y^n_{2,DPD}| W_1,\tf{S}^n)+n\epsilon_{n} 
 \end{eqnarray}
 \noindent where $(b)$ follows from Fano's inequality and~\eqref{l1u} by conditioning over $W_1$.
  
\noindent Next, we bound the sum-rate $R_1+R_2$  as follows.
\begin{align}
\label{term-3-mu}
& n(R_1+R_2)\notag\\&= {H}(W_1,W_2|z^n,\tf{S}^n)\notag \\
&\overset{(c)}{\le} I(W_1;y_1^n|z^n,\tf{S}^n)+ I(W_2;y_2^n|z^n, W_1,\tf{S}^n)+n\epsilon_{n}  \notag\\
&\le I(W_1;y_1^n|z^n,\tf{S}^n)+ I(W_2;y_1^n,y_2^n|z^n, W_1,\tf{S}^n)+n\epsilon_{n}  \notag\\
&= h(y_1^n|z^n,\tf{S}^n)-h(y_1^n|z^n, W_1, \tf{S}^n) +  I(W_2;y_1^n|z^n, W_1,\tf{S}^n)\notag\\&+  h(y_2^n|z^n,y_1^n, W_1,\tf{S}^n)- \underbrace{h(y_2^n|z^n,y_1^n, W_1,W_2,\tf{S}^n)}_{\geq no(\log(P))}+n\epsilon_{n}  \notag\\
&\overset{(d)}{\le} h(y_1^n|z^n,\tf{S}^n)  + h(y_2^n|z^n,y_1^n, W_1,\tf{S}^n) +n\epsilon_{n}  \notag\\
&\overset{(e)}{\le} h(z^n|\tf{S}^n) + h(y_{2}^n, y_1^n, z^n|  W_1,\tf{S}^n)-h( y_1^n, z^n| W_1,\tf{S}^n) +n\epsilon_{n}  \notag\\
&\overset{(f)}{\le} h(z^n|\tf{S}^n) + h(y_{2,PDD}^n| \tf{S}^n)\notag\\&+h( y_1^n, z^n, y_{2,DPD}^n| W_1,\tf{S}^n)-h( y_1^n, z^n| W_1,\tf{S}^n)  +n\epsilon_{n}  \notag\\
&\overset{(g)}{\le} h(z^n|\tf{S}^n) + h(y_{2,PDD}^n|\tf{S}^n)+\frac{1}{2}h( y_1^n, z^n, y_{2,DPD}^n| W_1,\tf{S}^n)+n\epsilon_{n}\notag\\
&{=} h(z^n|\tf{S}^n) + h(y_{2,PDD}^n|\tf{S}^n)+\frac{1}{2}h( y_1^n, y_{2,DPD}^n| W_1,\tf{S}^n)\notag\\&+\frac{1}{2}h( z^n| y_{2,DPD}^n, y_1^n, W_1,\tf{S}^n)+n\epsilon_{n} \notag\\
&{=} h(z^n|\tf{S}^n) + h(y_{2,PDD}^n|\tf{S}^n)+\frac{1}{2}h( y_1^n, y_{2,DPD}^n| W_1,\tf{S}^n)\notag\\&+\frac{1}{2}h( z_{PDD}^n| y_{2,DPD}^n, y_1^n, W_1,\tf{S}^n)\notag\\&+\frac{1}{2}\underbrace{h( z_{DPD}^n| y_{2,DPD}^n, z_{PDD}^n, y_1^n, W_1,\tf{S}^n)}_{\leq no(\log(P))}+n\epsilon_{n} \notag\\
&\overset{(h)}{\le} h(z^n|\tf{S}^n)+\frac{1}{2}h( z_{PDD}^n|\tf{S}^n) + h(y_{2,PDD}^n|\tf{S}^n)\notag\\&+\frac{1}{2}h( y_1^n, y_{2,DPD}^n| W_1,\tf{S}^n)+ no(\log(P))+n\epsilon_{n} \notag\\
&{\le} h(z_{DPD}^n|\tf{S}^n)+\frac{3}{2}h(z_{PDD}^n|\tf{S}^n) + h(y_{2,PDD}^n|\tf{S}^n)\notag\\&+\frac{1}{2}h( y_1^n, y_{2,DPD}^n| W_1,\tf{S}^n)+ no(\log(P)) +n\epsilon_{n}  
\end{align}
 \noindent where $\epsilon_n \rightarrow 0$ as $n \rightarrow \infty$; $(c)$ follows by following similar steps leading from~\eqref{sum-a} to~\eqref{sum-int}, $(d)$ follows  because $y_{2}^n$ can be obtained within bounded noise distortion form $(z^n, y_1^n, W_1,W_2,\tf{S}^n)$, $(e)$ follows by invoking~\eqref{l4u}, $(f)$ follows from the fact that conditioning reduces entropy, $(g)$ follows by~\eqref{l3u} with conditioning over $W_1$, and $(h)$ follow because $z_{DPD}^n$ can be obtained within bounded noise distortion form $(y_{2,DPD}^n, z_{PDD}^n, y_1^n, W_1,\tf{S}^n)$ and due to the fact that  conditioning reduces entropy.
    
Then, by applying Fourier-Motzkin elimination (for instance~\cite{NIT}) to eliminate  $h( y_1^n, y_{2,DPD}^n| W_1,\tf{S}^n)$ and $h( z^n_{DPD} | \tf{S}^n)$ from \eqref{term-1-mu}, \eqref{term-2-mu}, and \eqref{term-3-mu}, we get
\setlength{\arraycolsep}{0.2em}
\begin{eqnarray}
\label{final-sum-mu}
n(4R_1+ R_2)&\le& 3h( y_1^n | \tf{S}^n)+\frac{3}{2}h(z_{PDD}^n|\tf{S}^n) + h(y_{2,PDD}^n|\tf{S}^n)\notag\\&&+no(\log(P))+ n \epsilon'_{n}\notag\\
&\le& \big(3+\frac{3}{4} + \frac{1}{2}\big)n\log(P) +no(\log(P))+ n \epsilon'_{n}.
\end{eqnarray} 
Finally, dividing both sides of~\eqref{final-sum-mu} by $n\log(P)$ and taking $\lim {P \rightarrow \infty}$ and $\lim {n \rightarrow \infty} $, we get
\begin{eqnarray}
16d_1+4d_2 &\le& 17.
\end{eqnarray}
 
Due to the symmetry of the problem the proof of~\eqref{miso-mu-outer-2} follows along similar lines and is omitted for brevity. 

This concludes the proof.
\balance
\bibliographystyle{IEEEtran}
\bibliography{IEEEabrv,MISOBC-bib}
\nocite{isit16}
\end{document}